\documentclass[useAMS,usenatbib]{mnras}
\usepackage{float}
\usepackage{natbib}
\usepackage{amssymb,amsmath,amsfonts}
\usepackage{epsfig}
\usepackage{mathptmx}
\usepackage{longtable}
\usepackage{graphicx}
\usepackage{comment}
\usepackage{color}
\usepackage[caption=false]{subfig}
\usepackage{animate}
\usepackage{graphicx}
\usepackage{adjustbox}
\usepackage{subfig}
\usepackage[para]{bigfoot}
\makeatletter
\def\@fnsymbol#1{\ensuremath{\ifcase#1\or *\or \dagger\or \ddagger\or
   \mathsection\or \mathparagraph\or \|\or **\or \dagger\dagger
   \or \ddagger\ddagger \else\@ctrerr\fi}}
\makeatother
\DeclareNewFootnote[para]{default}[fnsymbol]
\MakeSortedPerPage{footnote}
\DeclareNewFootnote{L}
\MakeSorted{footnoteL}

\def\reff@jnl#1{{\rm#1\/}}

\def\aj{\reff@jnl{AJ}}                  
\def\araa{\reff@jnl{ARA\&A}}            
\def\apj{\reff@jnl{ApJ}}                        
\def\apjl{\reff@jnl{ApJ}}               
\def\apjs{\reff@jnl{ApJS}}              
\def\ao{\reff@jnl{Appl.Optics}}         
\def\apss{\reff@jnl{Ap\&SS}}            
\def\aap{\reff@jnl{A\&A}}               
\def\aapr{\reff@jnl{A\&A~Rev.}}         
\def\aaps{\reff@jnl{A\&AS}}             
\def\azh{\reff@jnl{AZh}}                        
\def\baas{\reff@jnl{BAAS}}              
\def\jrasc{\reff@jnl{JRASC}}            
\def\memras{\reff@jnl{MmRAS}}           
\def\mnras{\reff@jnl{MNRAS}}            
\def\pra{\reff@jnl{Phys. Rev. A}}         
\def\prb{\reff@jnl{Phys. Rev. B}}         
\def\prc{\reff@jnl{Phys. Rev. C}}         
\def\prd{\reff@jnl{Phys. Rev. D}}         
\def\prl{\reff@jnl{Phys. Rev. Lett}}      
\def\pasp{\reff@jnl{PASP}}              
\def\pasj{\reff@jnl{PASJ}}              
\def\qjras{\reff@jnl{QJRAS}}            
\def\skytel{\reff@jnl{S\&T}}            
\def\solphys{\reff@jnl{Solar~Phys.}}    
\def\sovast{\reff@jnl{Soviet~Ast.}}     
\def\ssr{\reff@jnl{Space~Sci.Rev.}}     
\def\zap{\reff@jnl{ZAp}}                        
\def\nat{\reff@jnl{Nature}}             

\def\p#1by#2{{\partial{#1} \over \partial{#2}}}
\def\pp#1by#2#3{{\partial^2{#1} \over \partial{#2}\partial{#3}}}
\def\d#1by#2{{{\rm d}{#1} \over {\rm d}{#2}}}
\def\dd#1by#2#3{{{\rm d}^2{#1} \over {\rm d}{#2}{\rm d}{#3}}}

\title[]{A plethora of diffuse steep spectrum radio sources in Abell 2034 revealed by LOFAR}
\label{firstpage}
\author[Shimwell et~al.]{T. W. Shimwell$^{1}$\thanks{E-mail: shimwell@strw.leidenuniv.nl}, 
  J. Luckin$^{1,2}$,
 M. Br\"{u}ggen$^{3}$,
 G. Brunetti$^{4}$,
  H. T. Intema$^{1,5}$, \newauthor
 M. S. Owers$^{6,7}$, 
 H. J. A. R\"{o}ttgering$^{1}$,
 A. Stroe$^{1,8}$, 
 R. J. van Weeren$^{9}$, 
 W. L. Williams$^{1,10}$, \newauthor
 R. Cassano$^{4}$,
 F. de Gasperin$^{1,3}$, 
  G. H. Heald$^{10,11}$\thanks{Current address: CSIRO Astronomy and Space Science, 26 Dick Perry Avenue, Kensington, Perth WA 6151, Australia},
 D. N. Hoang$^{1}$,
   M. J. Hardcastle$^{12}$,  \newauthor
  S. S. Sridhar$^{10,11}$,
 J. Sabater$^{13}$,
  P. N. Best$^{13}$,  
  A. Bonafede$^3$,
  K. T. Chy\.zy$^{14}$,
  T. A. En{\ss}lin$^{15}$,  \newauthor
  C. Ferrari$^{16}$,
  M. Haverkorn$^{17,1}$, 
  M. Hoeft$^{18}$,
     C. Horellou$^{19}$,  
   J. P. McKean$^{10,11}$,  \newauthor
   L. K. Morabito$^{1}$,
      E. Orr{\`u}$^{10,17}$, 
   R. Pizzo$^{10}$, 
   E. Retana-Montenegro$^{1}$,
   G. J. White$^{20,21}$ \\ \\
 $^1$ Leiden Observatory, Leiden University, PO Box 9513, NL-2300 RA Leiden, The Netherlands \\
  $^2$ Department of Physics, University of Maryland, Baltimore County, 1000 Hilltop Circle, Baltimore MD 21250, USA \\
 $^3$ Hamburger Sternwarte, Gojenbergsweg 112, 21029 Hamburg, Germany \\
 $^4$ INAF/Istituto di Radioastronomia, via Gobetti 101, I-40129 Bologna, Italy \\
 $^5$ National Radio Astronomy Observatory, 1003 Lopezville Road, Socorro, NM 87801-0387, USA \\
 $^6$ Department of Physics and Astronomy, Macquarie University, NSW 2109, Australia \\
 $^7$ Australia and Australian Astronomical Observatory PO Box 915, North Ryde NSW 1670, Australia \\
 $^8$ European Southern Observatory, Karl-Schwarzschild-Str. 2, 85748, Garching, Germany \\
 $^9$ Harvard-Smithsonian Center for Astrophysics, 60 Garden Street, Cambridge, MA 02138, USA \\
 $^{10}$ ASTRON, the Netherlands Institute for Radio Astronomy, Postbus 2, 7990 AA, Dwingeloo, The Netherlands \\
 $^{11}$ Kapteyn Astronomical Institute, PO Box 800, 9700 AV Groningen, The Netherlands \\
 $^{12}$ School of Physics, Astronomy and Mathematics, University of Hertfordshire, College Lane, Hatfield, Hertfordshire AL10 9AB, UK \\
 $^{13}$ SUPA, Institute for Astronomy, Royal Observatory, Blackford Hill, Edinburgh, EH9 3HJ, UK \\
 $^{14}$ Astronomical Observatory, Jagiellonian University, ul. Orla 171, 30-244 Krak\'ow, Poland  \\
 $^{15}$ MPI f. Astrophysik, Karl-Schwarzschild-Str. 1, 85741 Garching, Germany \\
 $^{16}$ Laboratoire Lagrange, UMR 7293, Universit\'e de Nice Sophia-Antipolis, CNRS, Observatoire de la C\^{o}te d'Azur, 06300 Nice, France \\
  $^{17}$ Department of Astrophysics/IMAPP, Radboud University, P.O. Box 9010, 6500 GL Nijmegen, The Netherlands\\
 $^{18}$ Th{\"u}ringer Landessternwarte, Sternwarte 5, 07778 Tautenburg, Germany \\
 $^{19}$ Department of Earth and Space Sciences, Chalmers University of Technology, Onsala Space Observatory, SE-439 92 Onsala, Sweden \\
$^{20}$  Department of Physical Sciences, The Open University, Milton Keynes MK7 6AA, England \\
$^{21}$ RAL Space, The Rutherford Appleton Laboratory, Chilton, Didcot, Oxfordshire OX11 0NL, England 
 \date{}}
\begin{document}
\maketitle
\begin{abstract}
\noindent
With Low-Frequency Array (LOFAR) observations, we have discovered a diverse assembly of steep spectrum emission that is apparently associated with the intra cluster medium (ICM) of the merging galaxy cluster Abell 2034. Such a rich variety of complex emission associated with the ICM has been observed in few other clusters. This not only indicates that Abell 2034 is a more interesting and complex system than previously thought but it also demonstrates the importance of sensitive and high-resolution, low-frequency observations. These observations can reveal emission from relativistic particles which have been accelerated to sufficient energy to produce observable emission or have had their high energy maintained by mechanisms in the ICM. The most prominent feature in our maps is a bright bulb of emission connected to two steep spectrum filamentary structures, the longest of which extends perpendicular to the merger axis for 0.5\,Mpc across the south of the cluster. The origin of these objects is unclear, with no shock detected in the X-ray images and no obvious connection with cluster galaxies or AGNs. We also find that the X-ray bright region of the cluster coincides with a giant radio halo with an irregular morphology and a very steep spectrum. In addition, the cluster hosts up to three possible radio relics, which are misaligned with the cluster X-ray emission. Finally, we have identified multiple regions of emission with a very steep spectral index that seem to be associated with either tailed radio galaxies or a shock.

\end{abstract}


\section{Introduction}
Diffuse synchrotron emission associated with ultra-relativistic particles and magnetic fields in the intra cluster medium (ICM) primarily consists of radio halos and radio relics (see \citealt{Ferrari_2008}, \citealt{Bruggen_2012}, \citealt{Feretti_2012} and \citealt{Brunetti_2014} for recent reviews). It is thought that radio halos are caused by  cluster-wide post-merger turbulence (see e.g. \citealt{Brunetti_2001} and \citealt{Petrosian_2001}), secondary electrons from proton-proton interactions (see e.g. \citealt{Dennison_1980} and \citealt{Blasi_1999}) or a combination of the two mechanisms (see \citealt{Brunetti_2005}, \citealt{Brunetti_2011} and \citealt{Pinzke_2015}), whereas radio relics are apparently associated with localised, post-merger shock-fronts (\citealt{Ensslin_1998}). However, the non-detection of gamma-ray emission by the Fermi satellite (see e.g. \citealt{Brunetti_2012}, \citealt{Zandanel_2014} and \citealt{Ackermann_2015}) disfavours a purely hadronic model for the origin of radio halos and challenges standard diffuse shock acceleration to explain radio relics (see e.g. \citealt{Vazza_2014} and \citealt{Brunetti_2014}). Additionally, the variety of the observed properties of cluster-scale radio emission is becoming increasingly difficult to describe within the current theoretical picture. For example: whilst in the `Sausage' cluster (CIZA J2242.8+5301) \cite{vanWeeren_2010} observe a textbook example of an arc-like radio relic related to a shock; in ZwCl 2341.1+000 \cite{Ogrean_2014} observe no X-ray shock at the position of a relic, and in  the Bullet cluster (\citealt{Shimwell_2015}), PLCKG287.0+32.9 (\citealt{Bonafede_2014}) and the Coma cluster (\citealt{Ensslin_1998}) an apparent link is seen between radio galaxies and radio relics. Furthermore, whilst radio halos are statistically detected in merging clusters (e.g. \citealt{Cassano_2013}), the role of the cluster mass and its dynamical state is difficult to disentangle with current observations (see \citealt{Cuciti_2015}). For example, in CL1821+643 \cite{Bonadefa_2014B} detect a giant radio halo in a cool core cluster with no obvious merging activity and \cite{Russell_2011} observe no diffuse radio emission in Abell 2146, which is a less massive cluster, but a clear merging system. Recently upgraded and new facilities have significantly improved sensitivity to diffuse radio emission from the ICM and are already beginning to reveal increasingly complex phenomena (e.g. \citealt{Owen_2014}) which may shed light on the connection between halos and relics and further challenge a univocal interpretation of these sources. One such instrument is the Low-Frequency Array (\citealt{vanHaarlem_2013}; LOFAR) which can produce deep, high-resolution, high fidelity, low frequency radio images.

In this publication, we present 118-166\,MHz LOFAR observations of Abell 2034, which is a galaxy cluster with known diffuse radio emission and good multi-wavelength datasets. Previous studies of this cluster have revealed that Abell 2034 is a massive cluster, with velocity measurements of the member galaxies by \cite{Owers_2014} implying a mass of $\rm{M}_{200}=1.1\pm0.4 \times 10^{15}M_{\odot}$ (where $\rm{r_{200}}$ is equal to 2.1\,Mpc) and Sunyeav Zel'dovich measurements by the \cite{Planck_2015} suggesting that $\rm{M}_{500}=4.19\pm0.4 \times 10^{14}M_{\odot}$ (where $\rm{r_{500}}$ is equal to 1.16\,Mpc). The luminosity of the cluster is also high ($\rm{L_{X,0.1-2.4 keV} = 3.51 \times 10^{44}}$ erg $\rm{s^{-1}}$; \citealt{Piffaretti_2011})  and the Sunyeav Zel'dovich signal is strong (the Arcminute Microkelvin Imager and Planck measured $Y_{500}$ to be 24$\pm$14 and 33$\pm$6 respectively; \citealt{AMI_Planck_2013} and \citealt{Perrott_2014}). Abell 2034 is a clear merging system with multiple mass concentrations (\citealt{Okabe_2008}, \citealt{vanWeeren_2011} and \citealt{Owers_2014}) and a patchy temperature distribution (\citealt{Kempner_2003} and \citealt{Owers_2014}). The most recent X-ray and optical study by \cite{Owers_2014} led to the discovery of a weak shock with Mach number $M=1.59^{+0.06}_{-0.07}$ at the Northern Edge of the cluster. \cite{Owers_2014} concluded that the merger axis is within $\sim23^\circ$ of the plane of the sky and that the two main components are now moving along the north-south direction $\sim$0.3\,Gyr after the core passage. Radio observations by \cite{Kempner_2001}, \cite{Rudnick_2009}, \cite{Giovannini_2009} and \cite{vanWeeren_2011} have revealed faint diffuse emission in the X-ray bright region of the cluster which is slightly enhanced in the vicinity of the X-ray detected shock. The nature of this emission is still disputed and it remains uncertain whether it should be classified as a radio halo, radio relic or both. Additional diffuse radio emission was identified by \cite{vanWeeren_2011} who classified a radio source far from the X-ray bright region in the western periphery of the cluster as a small radio relic. 

The primary aim of this low-frequency study of Abell 2034 is to search for previously unseen radio emission, to offer insights into the debate regarding the nature of the emission at the low Mach number shock front, and to assess the influence of the shock on that emission. We also aim to take advantage of the existing auxiliary data and the excellent sensitivity of our LOFAR data to steep-spectrum diffuse emission to examine whether there are interactions between the prominent radio sources in this cluster and the diffuse radio emission that appears to be associated with the ICM. 

Hereafter, we assume a concordance $\rm{\Lambda}$CDM cosmology, with $\rm{\Omega_{m}}$ = 0.3, $\rm{\Omega_\Lambda}$ = 0.7 and H$_{0}$ = 70 km\,s$^{-1}$Mpc$^{-1}$. At the redshift of the Abell 2034 ($z=0.1132$; \citealt{Owers_2014}) the luminosity distance is 526\,Mpc and 1$\arcsec$ corresponds to 2.06\,kpc. We use $S_{\nu} \propto {\nu}^\alpha$, where $S_{\nu}$ is the flux density and $\alpha$ is the spectral index. All coordinates are given in J2000.

\section{Observations and data reduction}
\subsection{LOFAR} \label{sec:lofar_reduction}

We used the LOFAR High Band Antenna (HBA) in HBA\_DUAL\_INNER mode to observe the galaxy cluster Abell 2034 on 2014 April 26 (observation ID L221519). Our observations were performed during the nighttime, centred on 15:10:10.8 +33:30:21.6,  and 8\,hours in duration. We observed the bright and well characterised calibrator source 3C295 for 25 minutes prior to the observation of our target. Both the target and the calibrator data were taken with frequency coverage from 118\,MHz to 190\,MHz. The data were recorded with 1\,s sampling  and 64 channels per 0.195\,MHz sub band. These high spectral and time resolution data were flagged for interference by the observatory with \textsc{AOFLAGGER} (\citealt{Offringa_2012}) before being averaged to 5\,s sampling and 4 channels per sub band. Only these averaged datasets were stored in the LOFAR archive and the raw data were deleted. The observations are summarised in Table \ref{LOFAR-obs}.

The calibration and imaging procedure mimics the facet calibration scheme that is thoroughly described by \cite{vanWeeren_2016a} and \cite{Williams_2016}. In this procedure, the calibration is performed in two stages: we first complete a direction independent calibration of the data before calibrating individual facets to remove time and position varying station beam errors and ionospheric effects. The procedure is briefly summarised below.

In the direction independent calibration step we initially averaged our 3C295 data to 10\,seconds and 2 channels per sub band, flag bad antenna stations (CS103HBA0 and CS501HBA1 leaving 43 core stations and 14 remote stations) and interference (all frequencies above 166\,MHz were removed completely). The XX and YY gains, as well as the rotation angle (to account for differential Faraday Rotation), were determined from calibrating the 3C295 data off a simple two point source model. From these calibration solutions the clock offsets for different antenna stations were then determined. The amplitude and the clock solutions were transferred to the target data before the target data were phase calibrated using a sky model generated from the VLA Low-Frequency Sky Survey (VLSSr; \citealt{Lane_2012}), Westerbork Northern Sky Survey (WENSS; \citealt{Rengelink_1997}) and the NRAO/VLA Sky Survey (NVSS; \citealt{Condon_1998}) -- see The LOFAR Imaging Cookbook\footnote{https://www.astron.nl/radio-observatory/lofar/lofar-imaging-cookbook} for  details. All VLSS sources within five degrees of the pointing centre with a flux greater than 1\,Jy were included in the phase calibration catalogue and these sources are matched with WENSS and NVSS sources to include the spectral properties of the sources in the phase calibration catalogue. 

In the direction dependent calibration step our aim is to accurately image Abell 2034, all the emission from which lies within $\approx$15$\arcmin$ of the pointing centre. To achieve this we must minimise the contamination from bright and nearby sources and correct for the ionospheric and beam errors in the direction of the cluster. The number of directions we must use in the calibration is far fewer than used by e.g. \cite{Williams_2016} because their aim was to create a wide-field (19\,deg$^2$) thermal noise limited image. We selected 13 directions away from the cluster and for each of these we performed direction dependent calibration to obtain good sky models and calibration solutions for each direction. The sources in each direction were progressively removed using the good sky models and direction dependent calibration solutions until just the emission from the region of the cluster was left in the data. As the emission from the cluster is complex we choose not to do direction dependent calibration directly on this object. Instead, we copied the calibration solutions obtained from a pair of nearby sources (FIRST J150909.6+340508 and FIRST J151007.0+335121) to the facet containing Abell 2034. These sources are separated from the cluster by just $\approx$30$\arcmin$ so the ionospheric effects should be similar to those at the pointing centre.

After our data for Abell 2034 were corrected for ionospheric and beam errors, the data were imaged in \textsc{CASA}\footnote{http://casa.nrao.edu/} using the multi-scale multi-frequency deconvolution algorithm (\citealt{Rau_2011}). When necessary, the resulting images can be corrected for primary beam attenuation in the image plane by dividing out the primary beam using a beam model obtained from \textbf{awimager} (\citealt{Tasse_2013}). However, the emission from Abell 2034 is within the 99\% power point of the primary beam and such a correction is not required for this study. In the imaging procedure, the images were \textsc{clean}ed to a depth of 0.5\,mJy/beam with deconvolution scale sizes equal to zero (point source), 3, 7, 9, 25, 60 and 150 times the pixel size, which is approximately a sixth of the synthesised beam (see Table \ref{LOFAR-obs}). The robust parameter within the \cite{Briggs_1995} weighting scheme was set to --0.75 to apply a more uniform weighting to visibilities across the $uv$-plane and decrease the sidelobes, and a Gaussian taper was applied to reduce the weight of the data from the long baselines to enhance the diffuse emission. This procedure was performed to create full-bandwidth (118-166\,MHz) Stokes I images. The data were imaged with various Gaussian tapers to examine how its observed structure changes as the synthesised beam FWHM is varied from varied from $5\arcsec$ to $40\arcsec$. An example of a medium-resolution 118-166\,MHz LOFAR image is shown in Figure \ref{fig:lofar_regions} and the properties of the different resolution images  are summarized in Table \ref{LOFAR-obs}.

To determine the uncertainty in our absolute flux calibration we compared fluxes measured from our narrow-band, low-resolution ($\approx25\arcsec$), wide-field, 151\,MHz primary beam corrected image of Abell 2034 to those recorded in the 151\,MHz LOFAR Multifrequency Snapshot Sky Survey preliminary catalogue (MSSS; \citealt{Heald_2015}). In total we compared the fluxes of 9 sources, which were selected because they are bright (signal to noise greater than 10 in both images), isolated (no other detected sources within 2$\arcmin$ in the Abell 2034 image), compact ($\leq30\arcsec$ in the Abell 2034 image)  and within 2$^\circ$ of Abell 2034. We found that the median ratio of peak fluxes was 0.9 and for integrated fluxes it was 1.0, with the MSSS peak flux being higher than that in the Abell 2034 image. The scatter in the flux ratios was low, with the standard deviation of the ratio of peak fluxes found to be 0.16 and the standard deviation of the total fluxes equal to 0.15. Throughout this publication we therefore adopt a 15\% absolute flux calibration error on our LOFAR images.

\begin{table}
\caption{A summary of our LOFAR observations towards Abell 2034.}
 \centering
 \label{LOFAR-obs}
\begin{tabular}{lccc}
\hline 
Coordinates & 15:10:10.8 33:30:21.6 \\
Amplitude calibrator & 3C295 \\
Frequency range & 118-190\,MHz (118-166\,MHz used)\\
Spectral resolution & 0.04875\,MHz \\
Time resolution & 5\,secs \\
On-source integration time & 8\,hours \\
Image RMS & 190$\mu$Jy/beam at 7$\arcsec\times5\arcsec$ resolution\\
	& 210$\mu$Jy/beam at 12$\arcsec\times10\arcsec$ resolution\\
	& 350$\mu$Jy/beam at 22$\arcsec\times19\arcsec$ resolution\\
	& 425$\mu$Jy/beam at 38$\arcsec\times37\arcsec$ resolution\\ \hline
 \end{tabular}
\end{table}

\subsection{Spectral index maps and integrated spectrum calculations}
\label{sec:spec_maps}

Abell 2034 was previously observed with the Giant Metrewave Radio Telescope (GMRT) at 610\,MHz and the Westerbork Synthesis Radio Telescope (WSRT) at 1.4\,GHz  (\citealt{vanWeeren_2011}). We use these existing WSRT and GMRT images, together with LOFAR images, to create high resolution 150\,MHz to 610\,MHz and low resolution 150\,MHz to 1.4\,GHz spectral index images. To measure the same spatial scales, for images used to create the 150\,MHz to 610\,MHz and the 150\,MHz to 1.4\,GHz spectral index  images we applied the observed GMRT and WSRT inner and outer $uv$-ranges to our LOFAR data during the imaging. These ranges were 102$\lambda$ to 52.8k$\lambda$ and 249$\lambda$ to 12.4k$\lambda$ for the GMRT and WSRT respectively. When imaging the LOFAR data we also weighted the visibilities such that the LOFAR images were of comparable resolution to the GMRT and WSRT images. For the LOFAR image used to create the 150\,MHz to 610\,MHz spectral index image we used a robust parameter of -0.75 which provided a resolution of $6.4\arcsec \times 4.4\arcsec$ which is comparable to the GMRT image resolution of $6.8\arcsec \times  4.3 \arcsec$. Whereas, for the LOFAR image used to create the 150\,MHz to 1.4\,GHz spectral index image we used a robust parameter of +0.4 to obtain a resolution of $26\arcsec \times 23\arcsec$, which was chosen because it is comparable to the WSRT image resolution of $30\arcsec \times 16 \arcsec$. Whilst the weighting we have used in our  imaging strategy does partially compensate for the different distribution of visibilities on the $uv$-plane of the GMRT, WSRT and LOFAR observations (see Figure \ref{fig:lofar_wsrt_gmrt_uv}), a more accurate comparison would be to match the $uv$-coverages as closely as possible or to use uniform weighting over the $uv$-plane. However, our approach was used because it provides better sensitivity to the large scale emission from Abell 2034. The pair of images used for each spectral index map were convolved with a Gaussian to produce images with exactly the same resolution which were regridded to have the same pixel size. The resulting 31$\arcsec$ 150\,MHz and 1.4\,GHz and the 7.8$\arcsec$ 150\,MHz and 610\,MHz images were then used to create spectral index ($\alpha$) maps where a power law spectral index was calculated for each pixel (3.3$\arcsec$). The error on the fit was calculated given the error on the GMRT or WSRT and LOFAR pixel values ($S_{1,err}$ and $S_{2,err}$) with

\begin{equation}
\alpha_{err} = \frac{1}{\rm{ln}\frac{\nu_1}{\nu_2}} \sqrt{\left(\frac{S_{1,err}}{S_1}\right)^2 + \left(\frac{S_{2,err}}{S_2}\right)^2},
\end{equation}
where $S_1$, $S_2$, $\nu_1$ and $\nu_2$ are the 150\,MHz and the 610\,MHz or 1.4\,GHz pixel values and frequency values, respectively. The standard deviation of pixels ($S_{2,err}$) that was measured on the GMRT image in the region of the cluster was 90$\mu$Jy/beam and in the high-resolution LOFAR image we measured a noise ($S_{1,err}$) of 185$\mu$Jy/beam. For the GMRT and the LOFAR images the noise is approximately constant in the region of interest. The standard deviation of pixels ($S_{2,err}$) that was measured close to the centre of the input WSRT image was 58$\mu$Jy/beam and the errors put into our spectral index analysis were 58$\mu$Jy/beam divided by the primary beam attenuation of $\rm{cos^6(c \times \nu \times r)}$ where c is a constant equal to 68 (see Section 5.1 of the Guide to Observations with the Westerbork Synthesis Radio Telescope revision 2 dated 2010/07/20), $\nu$ is the observing frequency in GHz and $r$ is the radius in degrees. The low-resolution LOFAR image has a constant noise level ($S_{1,err}$) of 350$\mu$Jy/beam in the region of interest. In the GMRT, WSRT and LOFAR images, we are sensitive to sufficiently large spatial scales (8$\arcmin$ at for the WSRT and 20$\arcmin$ for the GMRT assuming that the maximum recoverable scale is $0.6\times\lambda/D$) to not resolve out any of the cluster emission up to 1\,Mpc scales, but due to the significant differences in the LOFAR, GMRT and WSRT $uv$-coverages  (see Figure \ref{fig:lofar_wsrt_gmrt_uv}) we urge caution when interpreting the Mpc scale emission in our spectral index images and we stress that further multi-frequency observations are required to improve the spectral characterisation of this cluster. 

We calculated the uncertainty on our integrated flux density measurements for extended sources by adding in quadrature an absolute flux calibration error with the error on the integrated flux density derived from the image noise. We used absolute flux calibration errors of 5\% for WSRT (as in e.g. \citealt{vanWeeren_2012}), 10\% for GMRT (see e.g. \citealt{Chandra_2004})  and 15\% for LOFAR (see Section \ref{sec:lofar_reduction}).

\section{Results}

\begin{figure*}
   \centering
   \includegraphics[width=15cm]{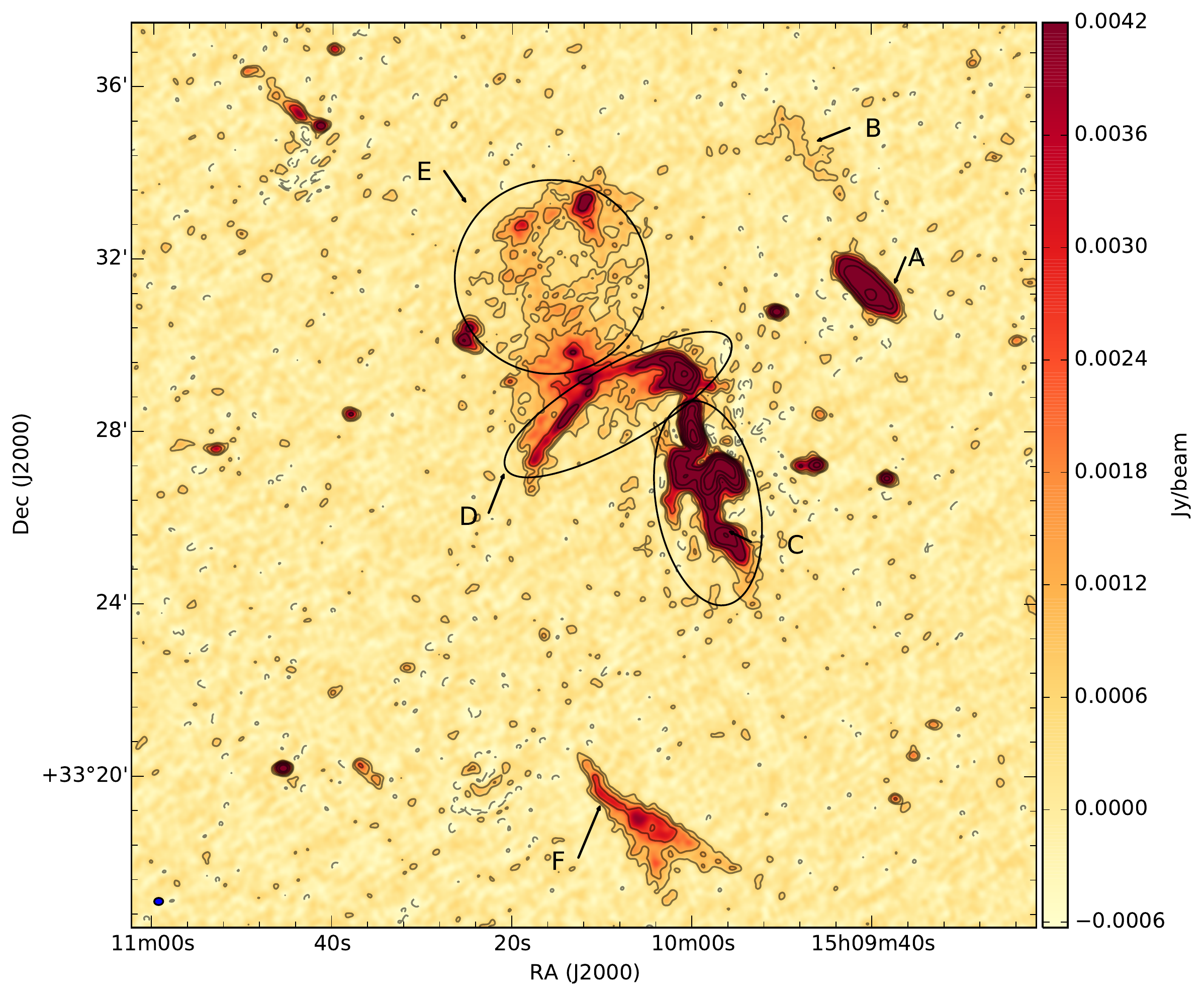}
   \caption{A medium-resolution (12$\arcsec \times$ 10$\arcsec$) 118-166\,MHz Stokes I image towards Abell 2034 with six regions of interest highlighted. The LOFAR contours show the $(1,2,4,...)\times 3 \times \sigma_{LOFAR,10\arcsec}$ levels where $\sigma_{LOFAR,10\arcsec} = 210\mu$Jy/beam.}
   \label{fig:lofar_regions}
\end{figure*}

\begin{figure*}
   \centering
   \includegraphics[height=4.61cm]{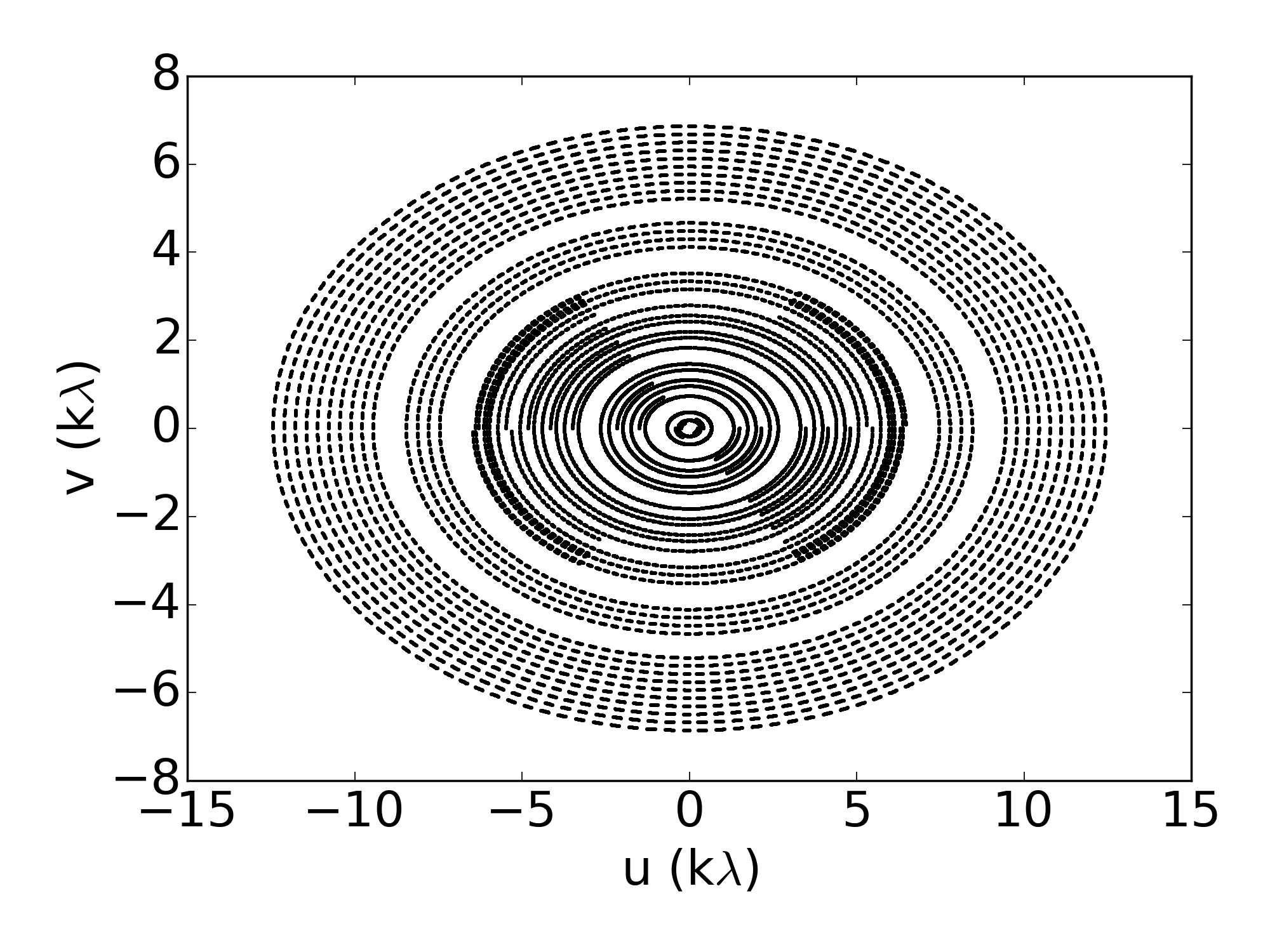}
   \hspace{-0.5cm}
   \includegraphics[height=4.61cm]{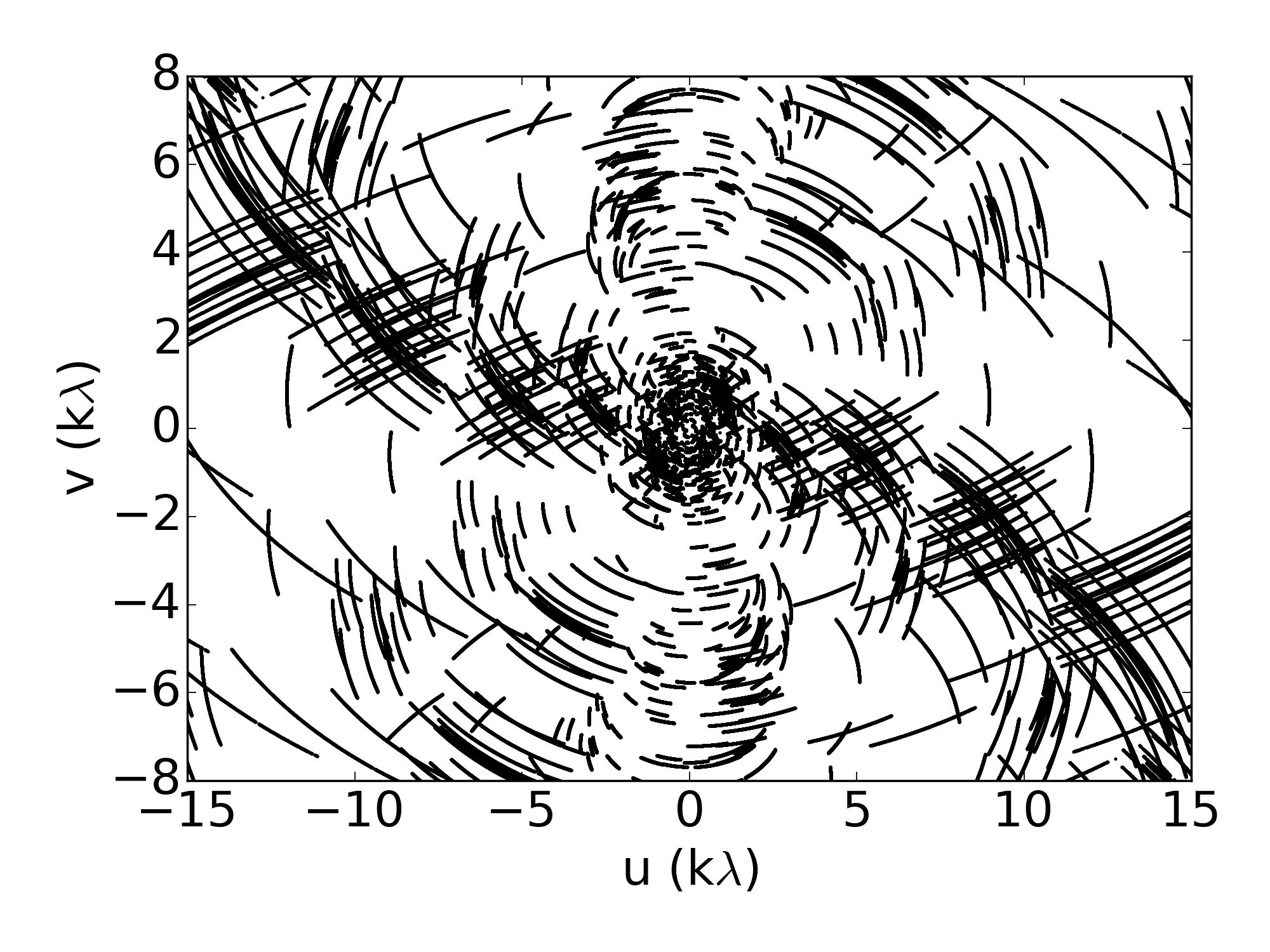}
   \hspace{-0.5cm}
   \includegraphics[height=4.61cm]{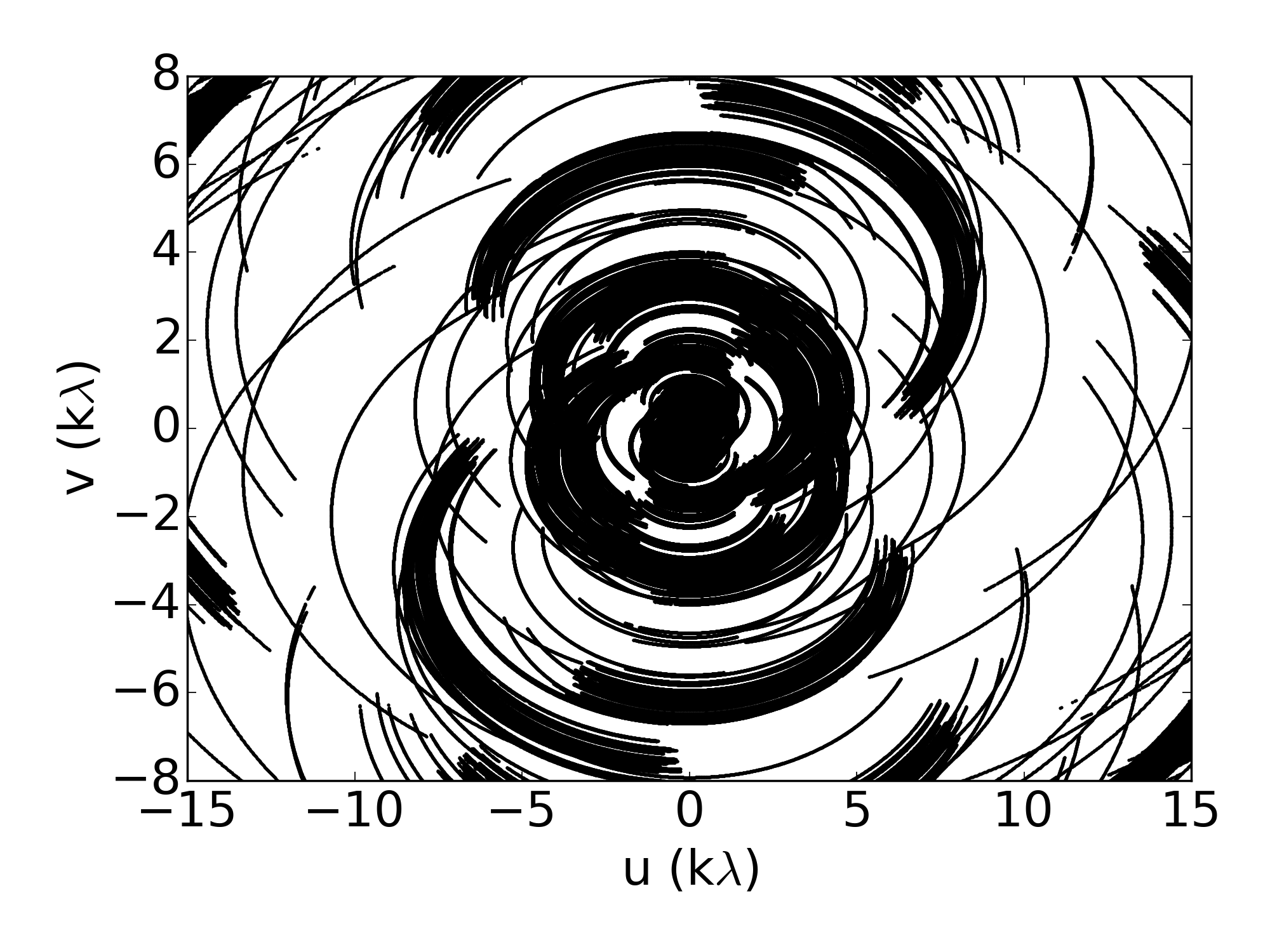}
   \vspace{-0.65cm}
   \caption{The monochromatic $uv$-coverage of the 1.4\,GHz WSRT (left), 610\,MHz GMRT (centre) and 150\,MHz LOFAR (right) observations of Abell 2034. This figure shows the entire WSRT $uv$-coverage but the GMRT and LOFAR coverage extends to 53k$\lambda$ and  60k$\lambda$ respectively. Here the monochromatic coverage has been presented for display purposes but the full bandwidth used in each observation (160\,MHz for WSRT, 32\,MHz for GMRT and 48\,MHz for LOFAR) provides considerable additional filling of the $uv$-plane.}
   \label{fig:lofar_wsrt_gmrt_uv}
\end{figure*}

\begin{figure*}
   \centering
   \includegraphics[width=15cm]{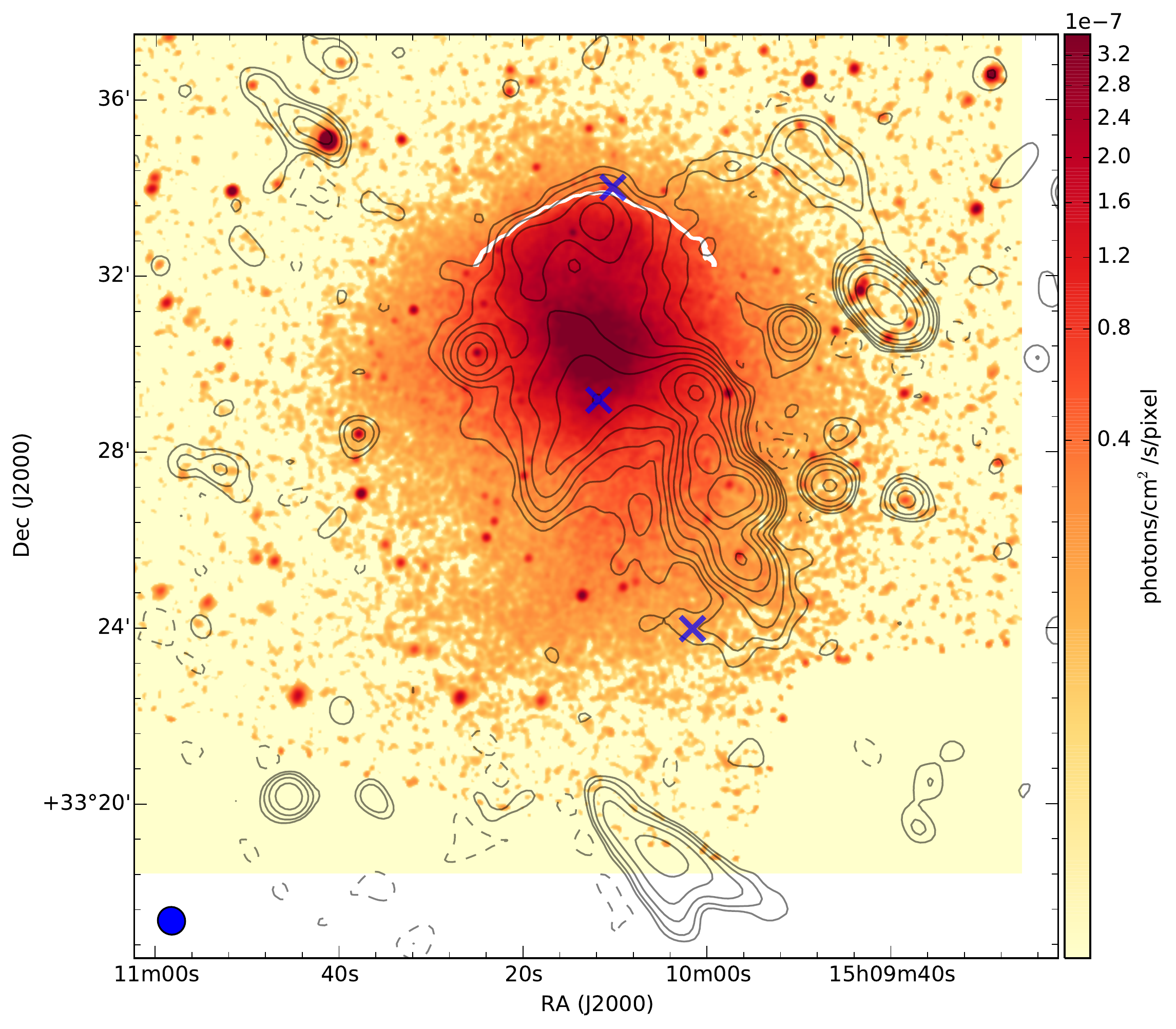}
   \caption{A low-resolution LOFAR image (38$\arcsec \times$ 37$\arcsec$) overlaid on an exposure-corrected, and background-subtracted \textit{Chandra} image in the 0.5--7.0 keV energy band with a total integration time of 250\,ks. The high-resolution \textit{Chandra} image (pixels of 1.968$\arcsec$) which was originally presented by \protect\cite{Owers_2014} has been smoothed with a Gaussian kernel with FWHM = 6$\arcsec$ and the thick white contour shows the X-ray detected shock front. The LOFAR contours show the $(1,2,4,...)\times 3 \times \sigma_{LOFAR,40\arcsec}$ levels where $\sigma_{LOFAR,40\arcsec} = 425\mu$Jy/beam. The two northern crosses show the prominent BCG of the two main merging clusters and the southern cross shows the position of a local peak in the galaxy surface density which may correspond to a third cluster involved in the merging (\protect\citealt{Owers_2014}).}
   \label{fig:lofar_xray}
\end{figure*}

\begin{figure*}
   \centering
   \includegraphics[width=15cm]{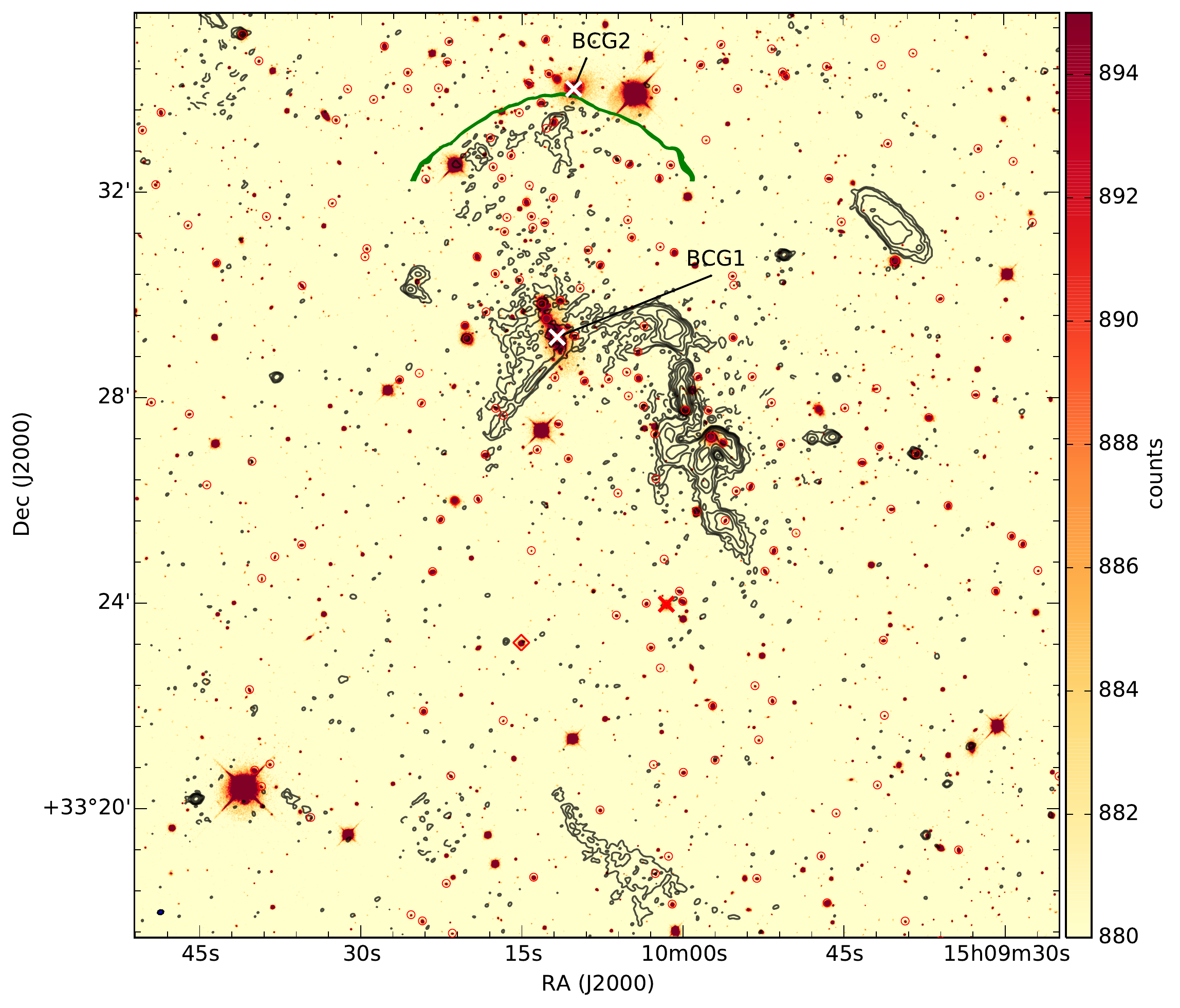}
    \includegraphics[height=5.5cm]{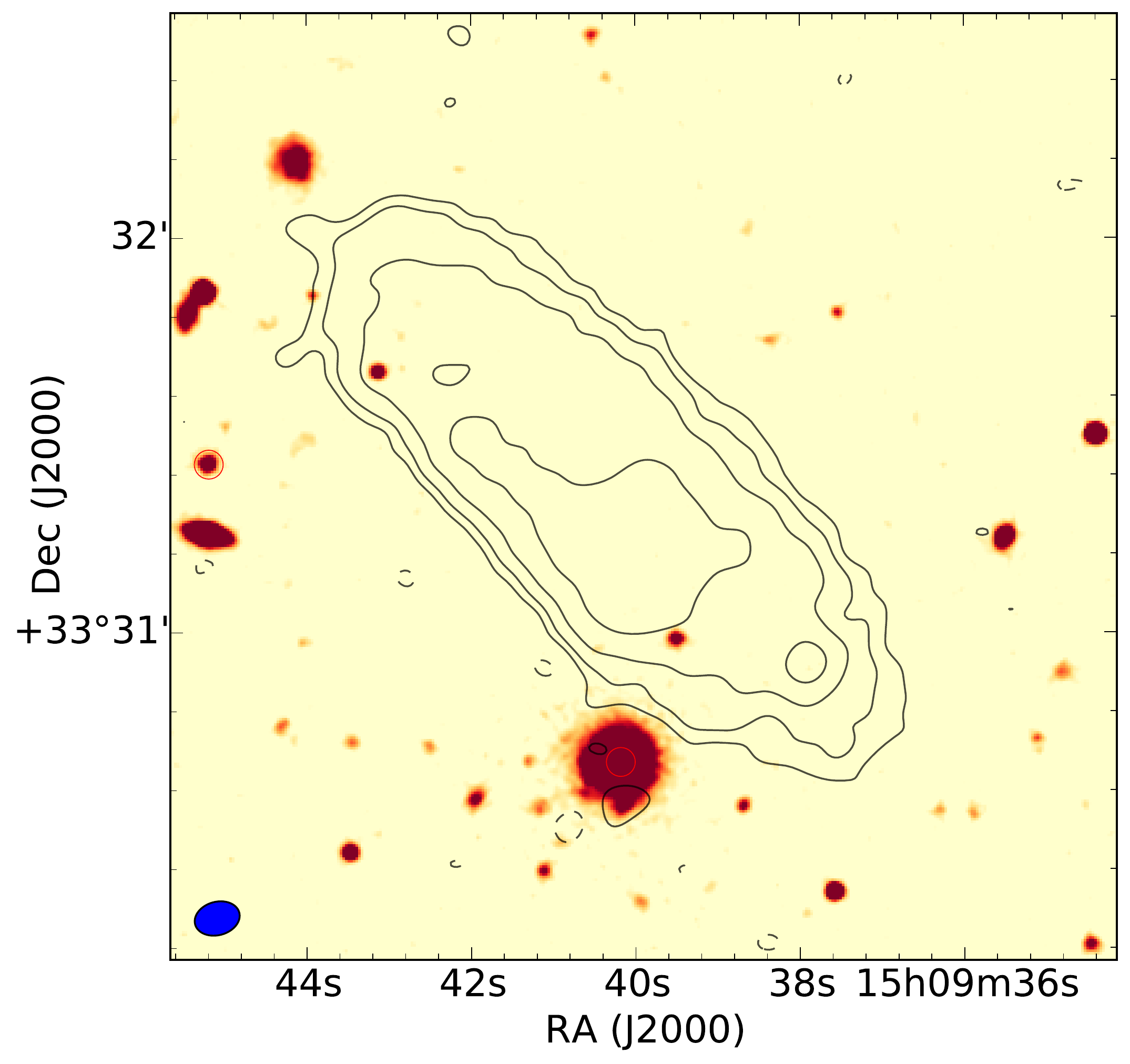}
    \includegraphics[height=5.5cm]{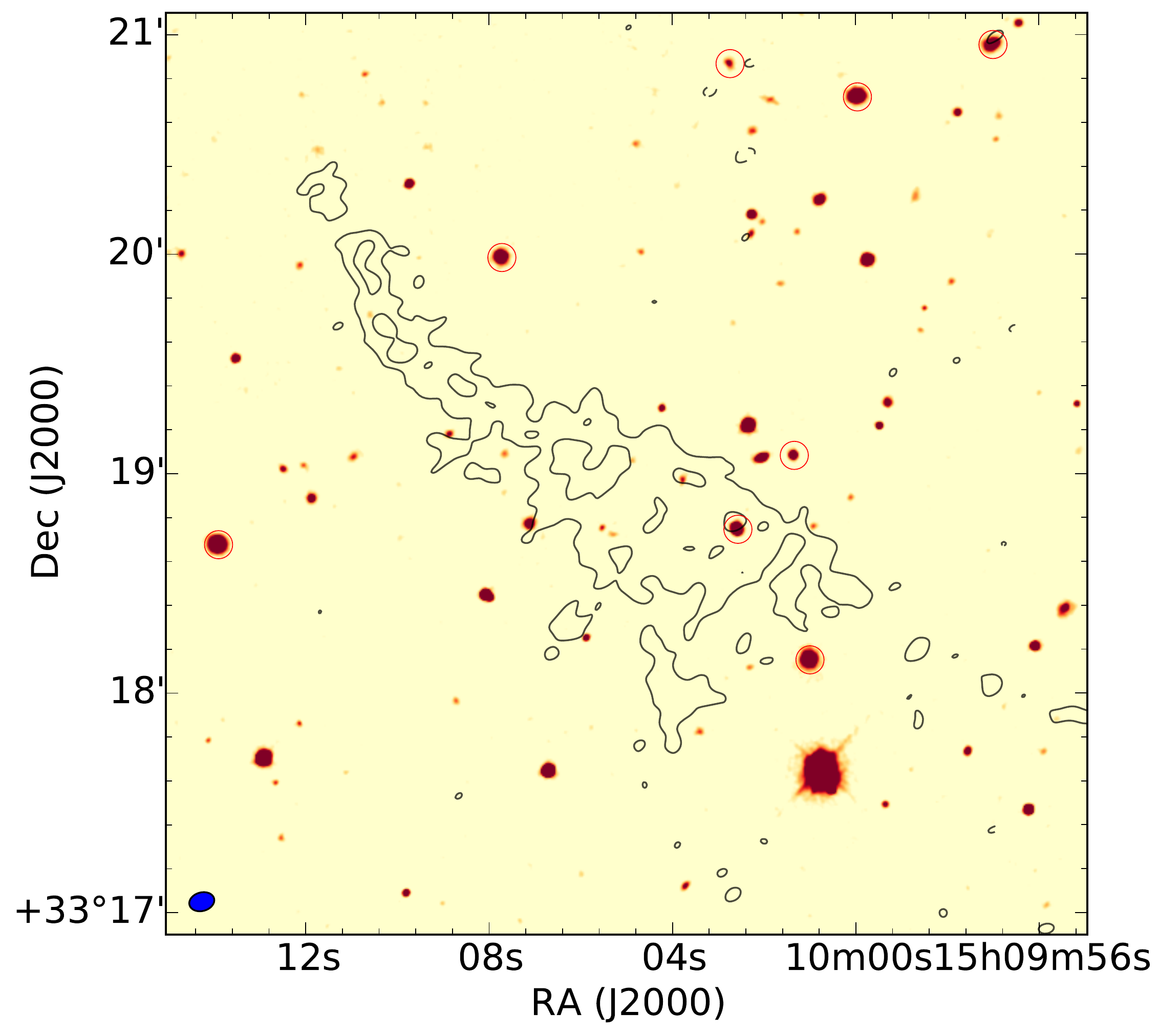}
    \hspace{-0.4cm}
    \includegraphics[height=5.5cm]{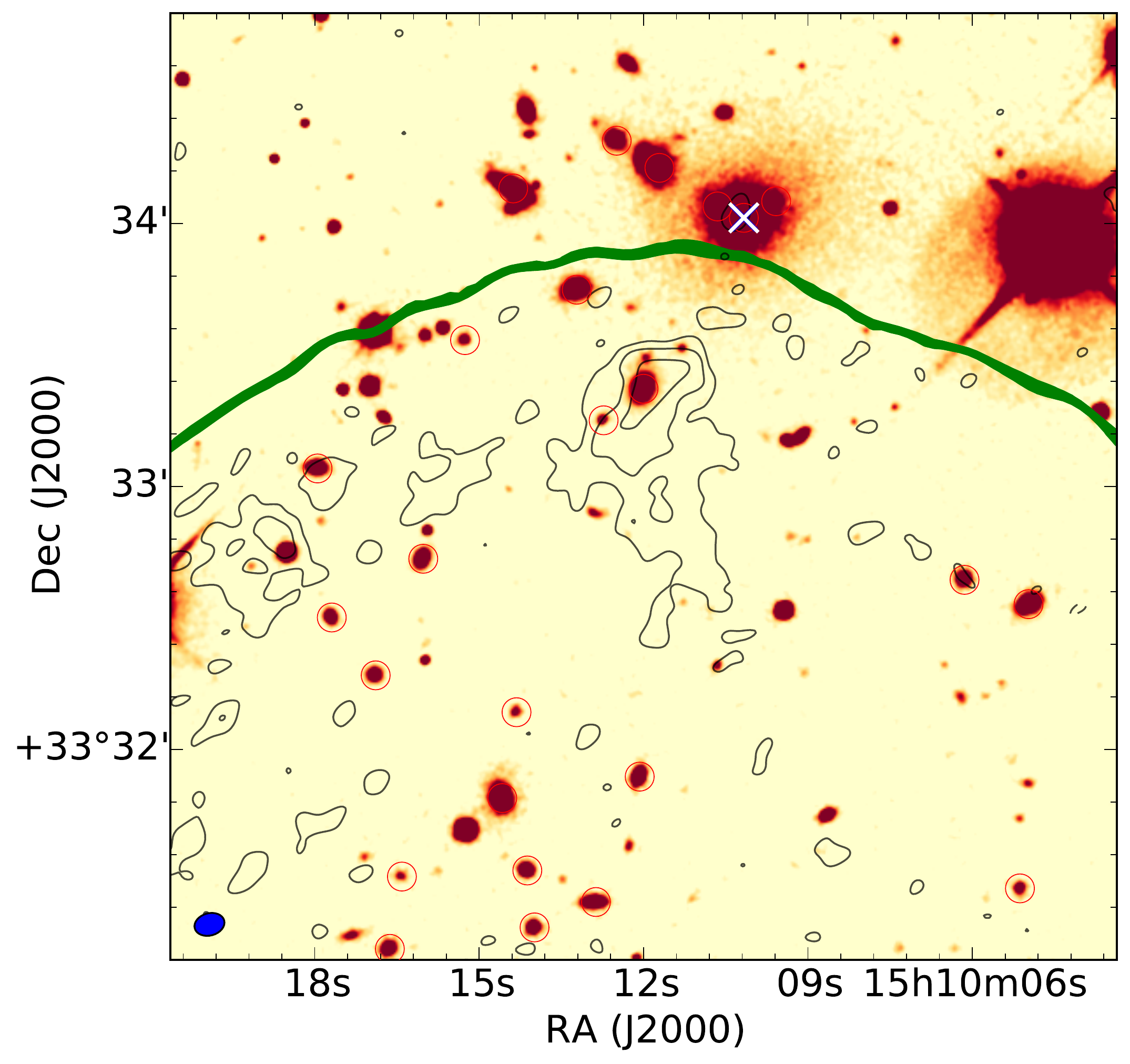}
   \caption{A high-resolution LOFAR image (7$\arcsec \times$ 5$\arcsec$) overlaid on an SDSS r-band image. The top panel shows the entire region of the cluster and the bottom panels show three regions of interest in more detail. The circles indicate the likely cluster members (based on their location in the redshift-radius phase space) and the two northern crosses show the prominent BCG of the two main merging clusters which we have also labelled BCG1 and BCG2 (see \citealt{Owers_2014}). A third, more southern, cross shows the position of a local peak in the galaxy surface density from \protect\cite{Owers_2014} which may correspond to the peak of a third cluster involved in the merging as was suggested by \protect\cite{Kempner_2003}. The red diamond shows a cluster member which is discussed in Section \ref{sec:relic_sources} and the thick green contour shows the X-ray detected shock front. The LOFAR contours show the $(1,2,4,...)\times 3 \times \sigma_{LOFAR,5\arcsec}$ levels where $\sigma_{LOFAR,5\arcsec} = 190\mu$Jy/beam.}
   \label{fig:lofar_optical}
\end{figure*}

\begin{figure*}
   \centering
   \includegraphics[height=7.3cm]{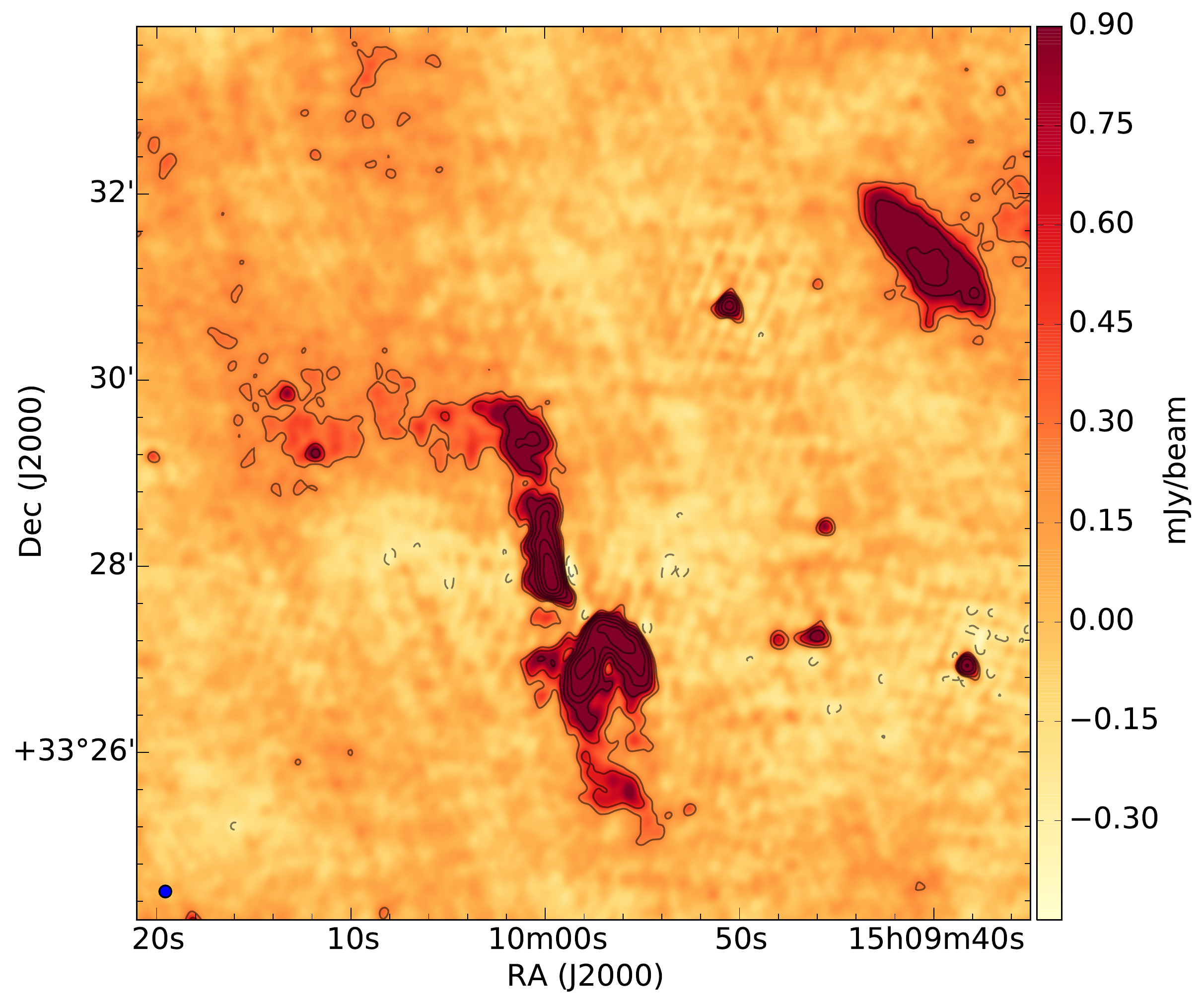}
   \includegraphics[height=7.3cm]{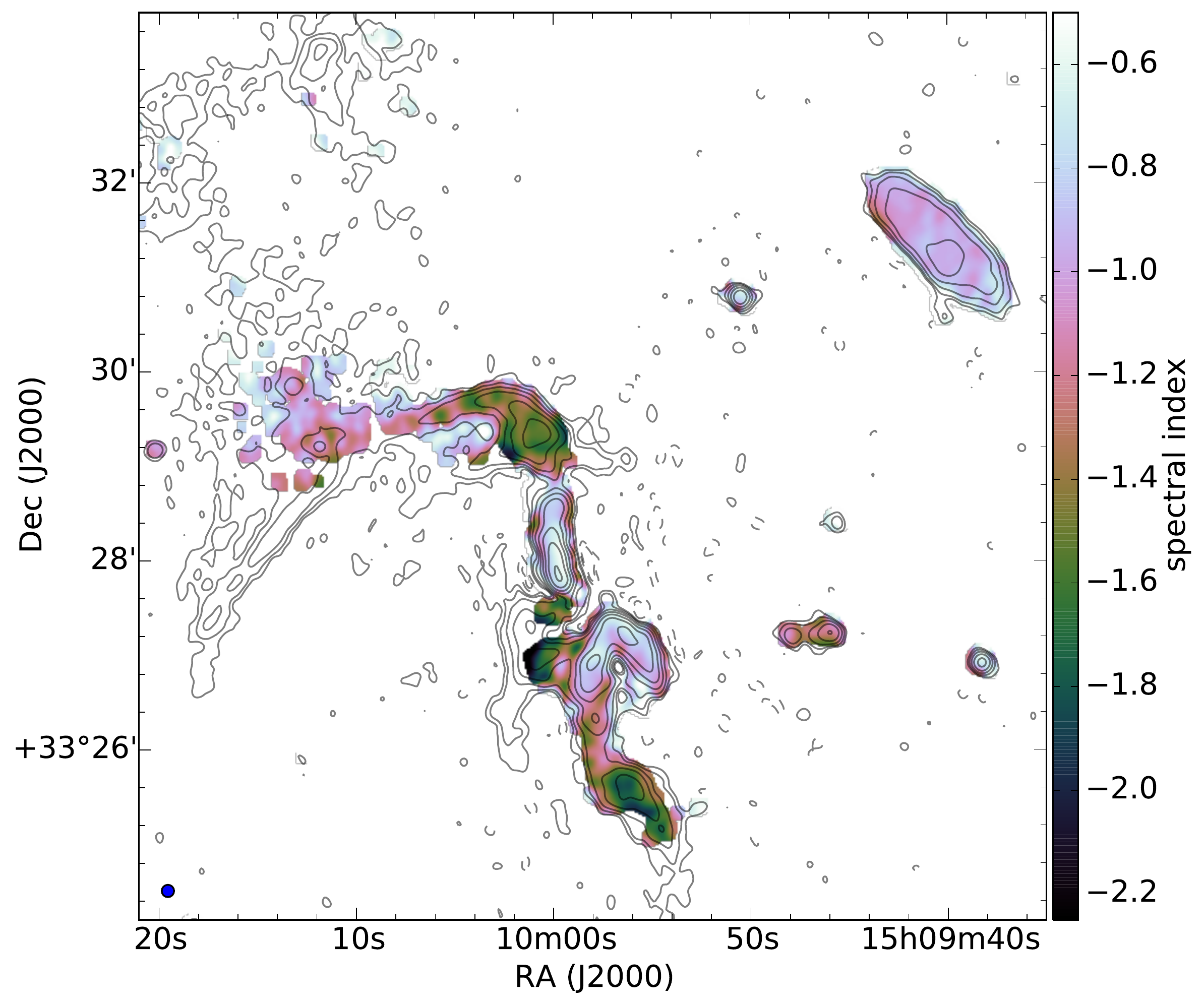}
   \caption{The 150\,MHz to 610\,MHz high-resolution ($7.8\arcsec \times 7.8\arcsec$) spectral properties of the emission from Abell 2034. Left: The 610\,GHz GMRT image with contours at $(1,2,4,...)\times 3 \times \sigma_{GMRT,8\arcsec}$ where $\sigma_{GMRT,8\arcsec} = 90\mu$Jy/beam. Right: The 150\.MHz to 610\,MHz high-resolution spectral index image where pixels with errors greater than 0.3 have been blanked.  The contours show a high-resolution ($7.8\arcsec \times 7.8\arcsec$)  150\,MHz LOFAR image with levels at $(1,2,4,...)\times 3 \times \sigma_{LOFAR,8\arcsec}$ where $\sigma_{LOFAR,8\arcsec} = 185\mu$Jy/beam. In both the 610\,MHz and 150\,MHz images the $uv$-range (102$\lambda$ to 52.8k$\lambda$) is the same and the maximum recoverable scale is $\approx20\arcmin$ which is sufficient to not resolve out any of the cluster emission but the resolution is high and the flux per pixel of the very low surface brightness structures is too low to be detected at high significance. The procedure for creating the GMRT and LOFAR images is outlined in Section \ref{sec:spec_maps}.}
   \label{fig:lofar_gmrt_spec}
\end{figure*}

\begin{figure*}
   \centering
   \includegraphics[height=7.3cm]{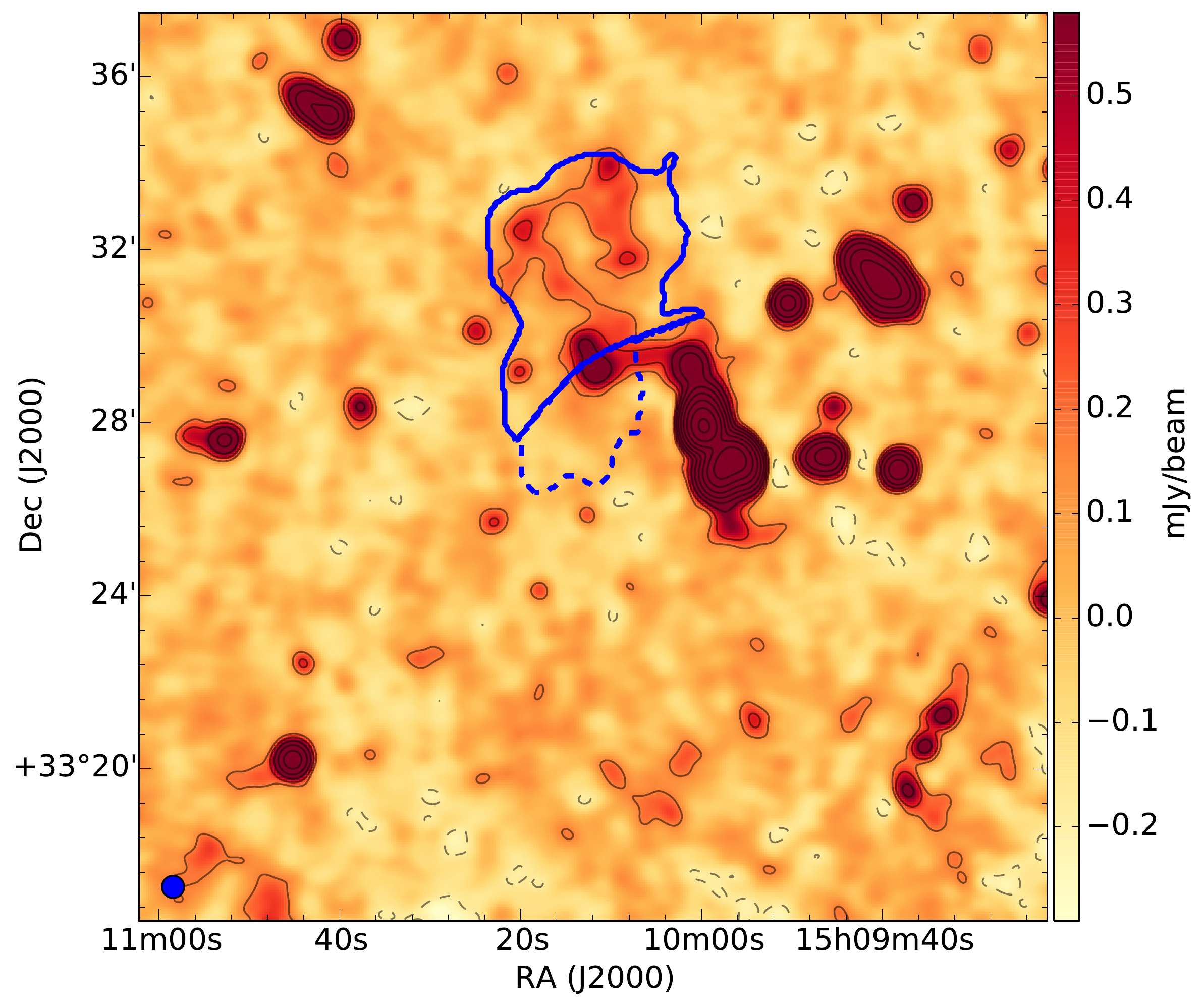}
   \includegraphics[height=7.3cm]{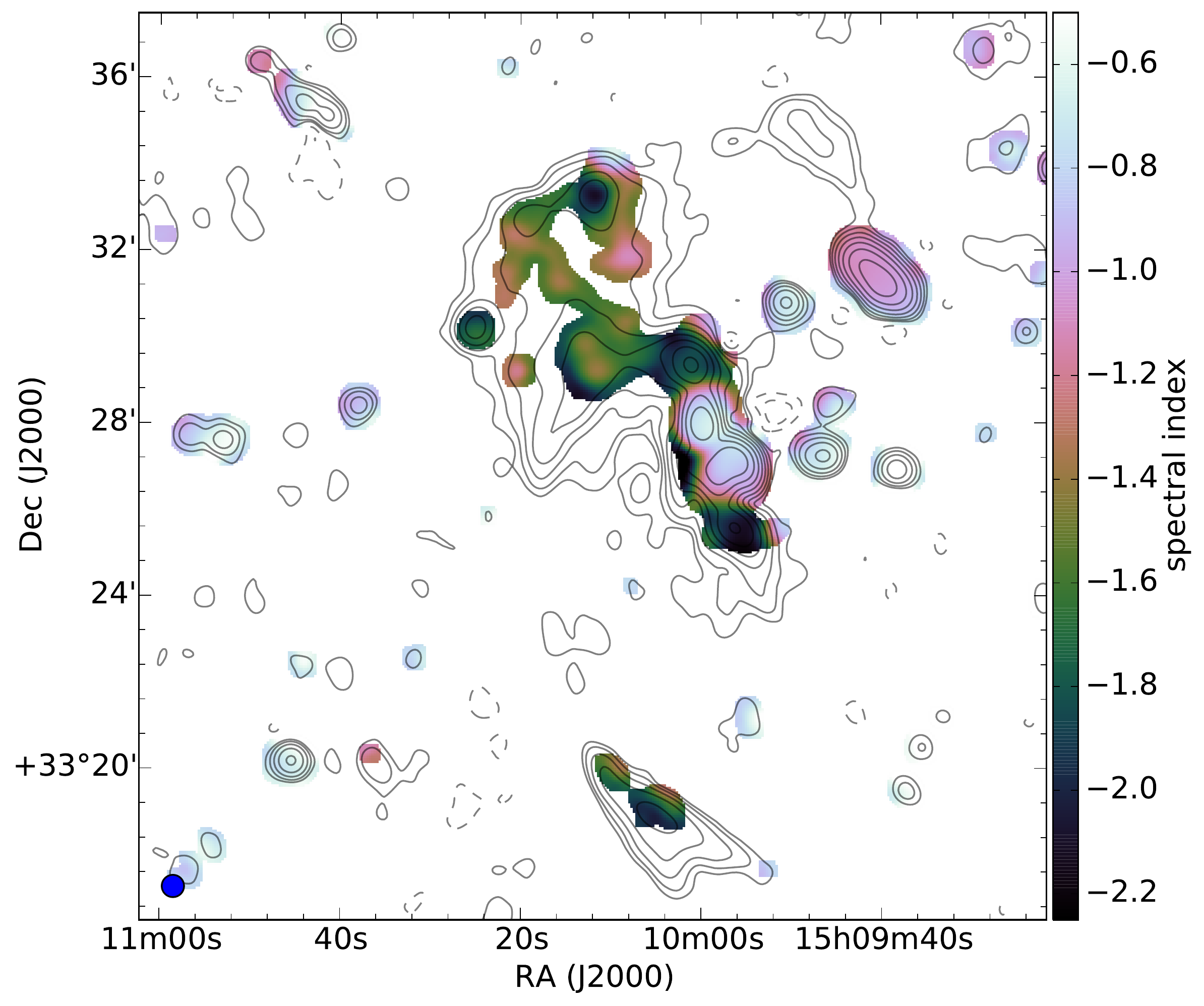}
   \caption{The 150\,MHz to 1.4\,GHz low-resolution ($31\arcsec \times 31\arcsec$) spectral properties of the emission from Abell 2034. Left: The 1.4\,GHz WSRT image with black contours showing the $(1,2,4,...)\times 3 \times \sigma_{WSRT,30\arcsec}$ levels where $\sigma_{WSRT,30\arcsec} = 58\mu$Jy/beam. The thick contours show the region used for integrated radio halo flux measurements (see Section \ref{sec:halo_integrated_flux}). Right: The 150\.MHz to 1.4\,GHz low-resolution spectral index image where pixels with errors greater than 0.3 have been blanked. The contours show a low-resolution ($31\arcsec \times 31\arcsec$) 150\,MHz LOFAR image with contour levels at $(1,2,4,...)\times 3 \times \sigma_{LOFAR,30\arcsec}$ where $\sigma_{LOFAR,30\arcsec} = 350\mu$Jy/beam. In both the 150\,MHz and 1.4\,GHz images the $uv$-range (249$\lambda$ to 12.4k$\lambda$) is the same and the maximum recoverable scale is $\approx8\arcmin$ which is sufficient to not resolve out any of the cluster emission. The procedure for creating the WSRT and LOFAR images is outlined in Section \ref{sec:spec_maps}.} 
   \label{fig:lofar_wsrt_spec}
\end{figure*}

In Figure \ref{fig:lofar_xray} we present a LOFAR 118-166\,MHz image of Abell 2034 which is at a resolution of $38\arcsec \times 37\arcsec$  and reaches a noise level of 425$\mu$Jy/beam. We have overlaid this image on the \textit{Chandra} X-ray image from \cite{Owers_2014} to show the correspondence between the X-ray and radio emission. A higher-resolution ($7\arcsec \times 5\arcsec$) LOFAR image of the same region with a noise level of 190$\mu$Jy/beam is presented in Figure \ref{fig:lofar_optical} and overlaid on an SDSS r-band image with known cluster members that were identified in the redshift-radius phase-space by \cite{Owers_2014} highlighted. We present the high-resolution (7.8$\arcsec$) 150\,MHz--610\,MHz and low-resolution 150\,MHz--1.4\,GHz spectral index images in Figures \ref{fig:lofar_gmrt_spec} and \ref{fig:lofar_wsrt_spec} respectively. The low-frequency LOFAR images are significantly more sensitive to the steep spectrum diffuse radio emission from the ICM than the existing 610\,MHz GMRT and 1.4\,GHz WSRT images which were presented \cite{vanWeeren_2011}. These deeper images have allowed us to discover new complex steep spectrum emission and characterise known objects in greater detail. In the LOFAR image, there are six main regions of diffuse emission associated with  Abell 2034 which we have labelled in Figure \ref{fig:lofar_regions}.

\subsection{Diffuse source A}

Using GMRT and WSRT radio images together with optical datasets,  \cite{vanWeeren_2011} classified source {\it{A}} (centred at 15:09:40 +33:31:20) as a small radio relic but also noted that the emission could be related to plasma from a nearby radio galaxy and that a more precise classification could be obtained with polarimetric and resolved spectral index measurements. \cite{vanWeeren_2011} measured source {\it{A}} to have a size of 220\,kpc $\times$ \,75\,kpc, a spectral index $\approx$--1.2 and an integrated 1.382\,GHz flux of $24\pm2$\,mJy, which, at the redshift of Abell 2034, corresponds to a 1.4\,GHz radio power of $8.0^{+0.7}_{-0.6}\times 10^{23}$\,W\,Hz$^{-1}$ with our adopted cosmology. From the low-resolution  WSRT and LOFAR images presented in Figure \ref{fig:lofar_wsrt_spec}, we measure the integrated flux density of source {\it{A}}  in the same region on both maps (defined as being above $4\times \sigma_{LOFAR,30\arcsec}$) to be $S_{150\,MHz} = 310\pm47$\,mJy and $S_{1.4\,GHz} = 28.5\pm1.5$\,mJy. This gives a spectral index measurement of $\alpha_{150-1400} = -1.07\pm0.07$. Similarly, from the high-resolution GMRT and LOFAR images (Figure \ref{fig:lofar_gmrt_spec}), we measure the integrated flux density (in the region with a 150\,MHz flux density above $4\times \sigma_{LOFAR,8\arcsec}$) to be $S_{150\,MHz} = 310\pm47$\,mJy and $S_{610\,MHz} = 76.8\pm7.7$\,mJy which corresponds to $\alpha_{150-610} = -0.99\pm0.13$. Both the minor and major axis of the source are well resolved in the high-resolution 150MHz to 610MHz spectral index map (Figure \ref{fig:lofar_gmrt_spec}), which reveals a fairly uniform spectral index distribution with small variations giving it a slightly patchy appearance. The typical spectral index values range from --0.9 to --1.0 and we observe no obvious trend in the direction towards the cluster centre as may be expected for a radio relic. The spectral index measurements in the the south western region are flatter ($\approx$--0.7) but this area is likely contaminated by a faint point-like source which is visible in our highest resolution images (see Figure \ref{fig:lofar_optical}). The 150\,MHz to 1.4\,GHz spectral index map (see Figure \ref{fig:lofar_wsrt_spec}) is in broad agreement with the higher resolution 150\,MHz to 610\,MHz spectral index map, but due to the low-resolution of the image and the contamination from the compact source in the south western region, it is difficult to accurately assess the resolved 150\,MHz to 1.4\,GHz spectral properties of source {\it{A}}.

\subsection{Diffuse source B}

The diffuse and very faint source {\it{B}} on the periphery of the cluster was previously undetected and is best displayed in our  lower resolution LOFAR images (see Figure \ref{fig:lofar_xray} or \ref{fig:lofar_wsrt_spec}). The brightest part of the structure is located around 15:09:48 +33:34:43 and has a peak flux density exceeding 15 times the noise in our low-resolution images; it is orientated in a direction similar to source {\it{A}} and runs approximately parallel to the closest X-ray brightness edge. From our low-resolution image in Figure  \ref{fig:lofar_wsrt_spec}  source {\it{B}} has an integrated 150\,MHz flux of $32.0\pm4.9$\,mJy (within the 4$\sigma_{LOFAR,30\arcsec}$ contour) and its non-detection within the same region on the WSRT image of the same resolution implies a 3$\sigma$ upper limit on the 1.4\,GHz flux of 0.6mJy allowing us to constrain its spectral index to be steeper than --1.8. Although there are several cluster members within the region of diffuse emission (see Figure \ref{fig:lofar_optical}), there is no obvious optical counterpart and no compact radio structure visible in our high-resolution low frequency images. At three times the noise level of our low-resolution images we detect faint bridges of emission between this object, the northern edge of the candidate radio relic (source {\it{A}}) and the north-western edge of the radio halo (source {\it{E}}). Because of the low significance of the detection of a faint bridge, its existence will need to be confirmed by deeper observations. In any case, based on the morphology, the lack of an optical counterpart and its location in the cluster periphery, the properties of this faint  emission in region {\it{B}} are similar to those expected for a radio relic. We stress that spectral and polarimetric observations are necessary to  robustly classify this source. 

\subsection{The complex regions C and D}

\begin{figure}
   \centering
   \includegraphics[width=8cm]{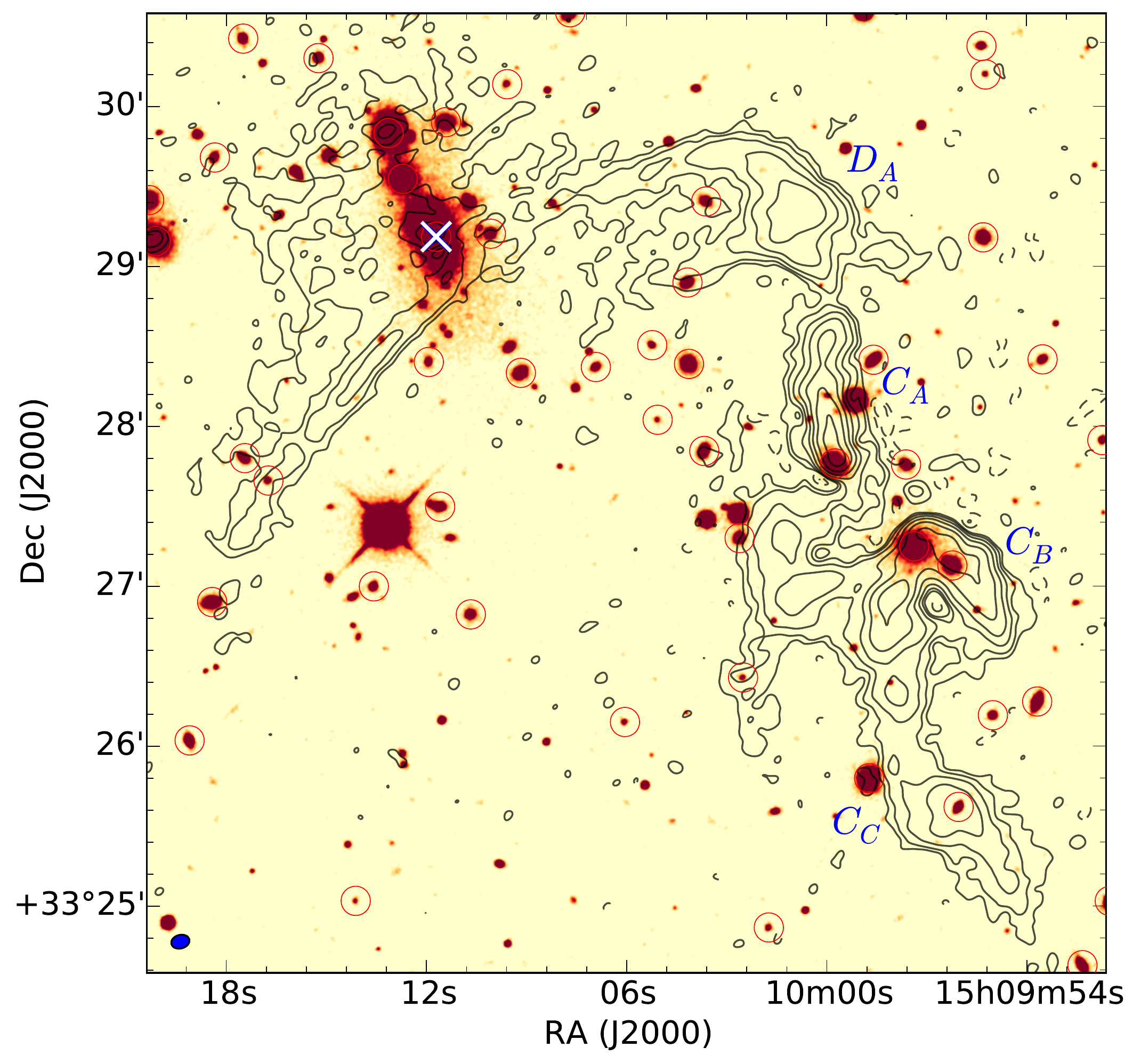}
   \caption{A high-resolution LOFAR image (7$\arcsec \times$ 5$\arcsec$) overlaid on an SDSS r-band image of region {\it{C}} and {\it{D}} of Abell 2034. Similarly to Figure \ref{fig:lofar_optical} the BCG is shown with a cross and the circles indicate likely cluster members. The tailed radio galaxies in region {\it{C}} and the bulb of emission in region {\it{D}} are labelled. The LOFAR contours show the $(1,2,4,...)\times 3 \times \sigma_{LOFAR,5\arcsec}$ levels where $\sigma_{LOFAR,5\arcsec} = 190\mu$Jy/beam.}
   \label{fig:lofar_headtail}
\end{figure}

To the south-west of the cluster is region C, which we show enlarged at high-resolution in Figure \ref{fig:lofar_headtail}.  In this region there are two prominent tailed radio galaxies with optical counterparts at the cluster redshift, which we refer to as $\rm{C_{A}}$ and $\rm{C_{B}}$. Around these galaxies we detect very steep spectrum diffuse emission (see Figures \ref{fig:lofar_gmrt_spec} and \ref{fig:lofar_wsrt_spec}), notably the emission labelled $\rm{C_{C}}$ which connects to the eastern tail of $\rm{C_{B}}$ and extends southwards with a peak of emission coincident with another cluster member; and the previously unseen emission connected to $\rm{C_{A}}$ at the position of its optical counterpart which extends eastwards and then south towards the head of $\rm{C_{B}}$.

In region {\it{D}}, we observe a very unusual feature, with a bright bulb of emission at 15:10:00 +33:29:20 with no obvious optical counterpart in the SDSS R band image (Figure \ref{fig:lofar_headtail}), that we refer to as $\rm{D_{A}}$. To the north eastern edge of $\rm{D_{A}}$ we observe a filament of emission that appears to emerge from the bulb of emission. This filament extends eastwards for $\sim$0.5\,Mpc (245$\arcsec$) with a slight break in its direction after 195\,kpc (95$\arcsec$) just south of the brightest cluster galaxy (BCG) which is itself a faint radio source. There is additional emission protruding from the south eastern edge of $\rm{D_{A}}$ which extends eastwards and has similar morphology to the filament, but its extension is only $\approx$ 60\,kpc (30$\arcsec$). It is unclear whether $\rm{D_{A}}$ and these filamentary structures are related or whether their apparent connection is a chance alignment. However, there is no obvious optical counterpart to either filament. 

\subsection{The diffuse region E}
\label{sec:halo_integrated_flux}
Both our high- and low-resolution images (Figures \ref{fig:lofar_xray} and \ref{fig:lofar_optical}) indicate that the X-ray bright region of the cluster (region {\it{E}}) hosts diffuse radio emission, which should most likely be classified as a radio halo. Measuring the integrated flux of the radio halo is challenging due to the contamination from radio sources and the difficulty in properly disentangling it from the bright emission in region {\it{D}}. However, we estimate the total emission from the low resolution WSRT and LOFAR images we presented in Figure \ref{fig:lofar_wsrt_spec} by integrating the flux in the area north of region {\it{D}} which has LOFAR emission exceeding 4$\sigma_{LOFAR,30\arcsec}$, this region is shown by the solid thick contour in the left panel of Figure \ref{fig:lofar_wsrt_spec}.  We measure the integrated flux in this region to be $S_{150\,MHz} = 416\pm62$\,mJy and $S_{1.4\,GHz} = 8.5\pm0.6$\,mJy which implies a spectral index of $\alpha \approx -1.7\pm0.1$. For comparison, we  note that \cite{Giovannini_2009} observed Abell 2034 at 1.4\,GHz using the VLA C and D configurations (in these data the shortest baseline was 35\,m whereas the shortest baseline used during the WSRT observation was 52\,m)  and measured the flux of the radio halo to be of 13.6$\pm$1.0\,mJy. Whilst we do not have the exact area used for this integrated flux measurement we approximate it with the total area within  the solid and dashed contours displayed in the left panel of Figure \ref{fig:lofar_wsrt_spec} which encompass the emission seen by \cite{Giovannini_2009}. We estimate the 150\,MHz integrated flux density of the radio halo in this region by integrating all the emission and subtracting the contaminating flux from BCG1 (a peak flux of 4.1\,mJy/beam at 150\,MHz) and the filament (which we estimate from our high resolution images to be approximately 100\,mJy). We calculate a spectral index of $\alpha \approx -1.6\pm0.1$ between our integrated flux measurement and the \cite{Giovannini_2009} measurement.

\begin{figure*}
   \centering
   \includegraphics[width=5.7cm]{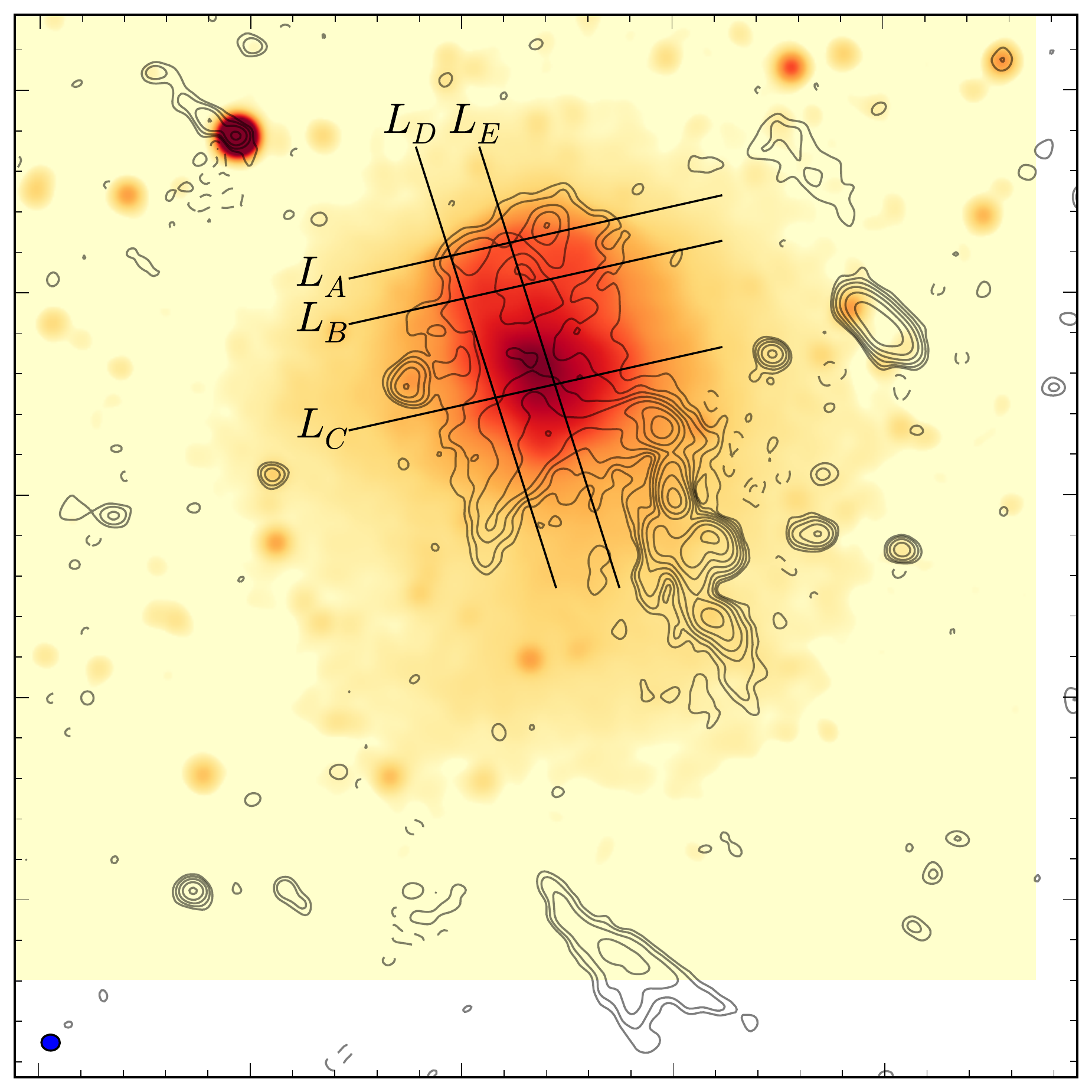}
   \includegraphics[height=4.3cm]{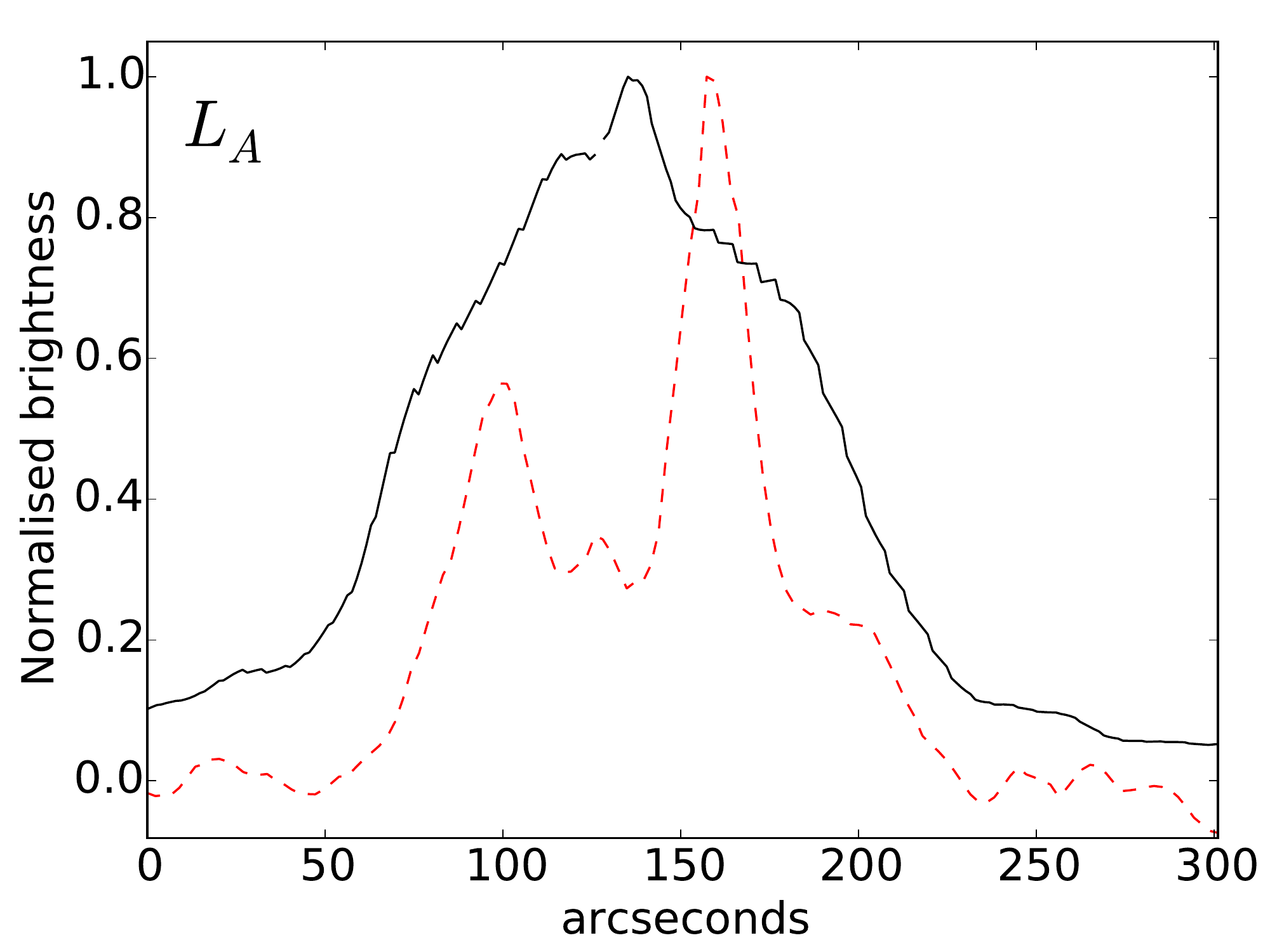}
   \includegraphics[height=4.3cm]{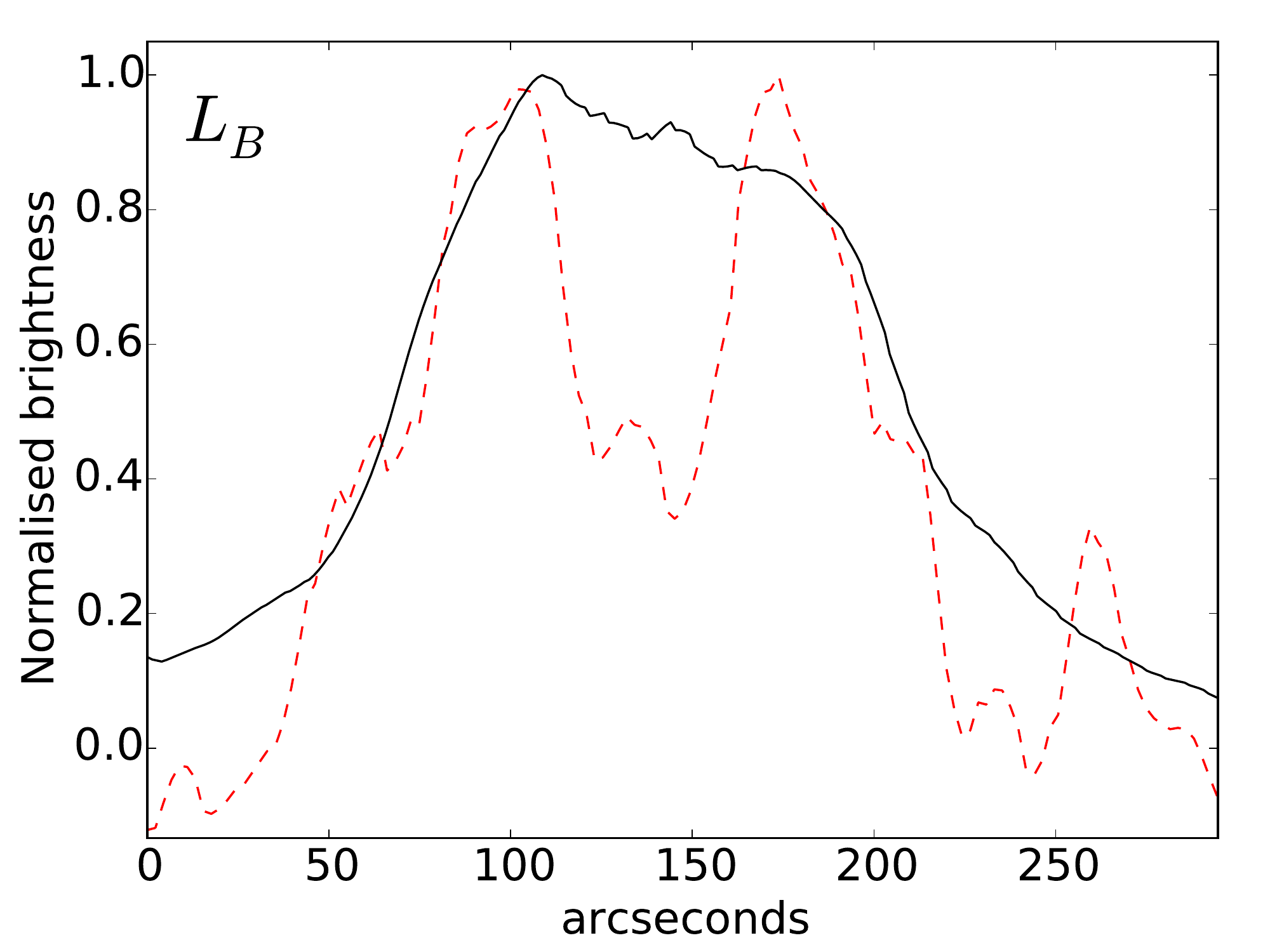}
   \includegraphics[height=4.3cm]{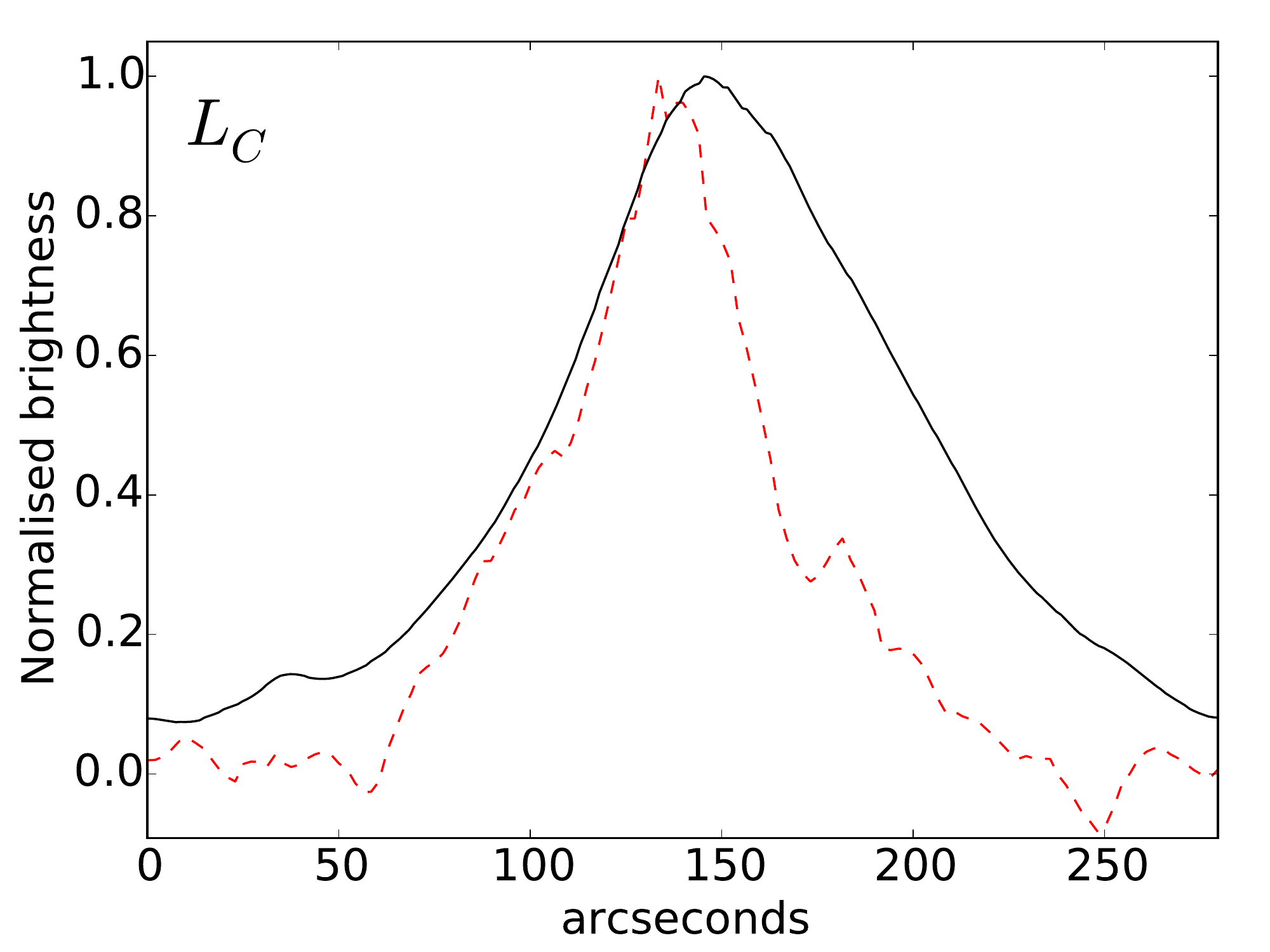}
   \includegraphics[height=4.3cm]{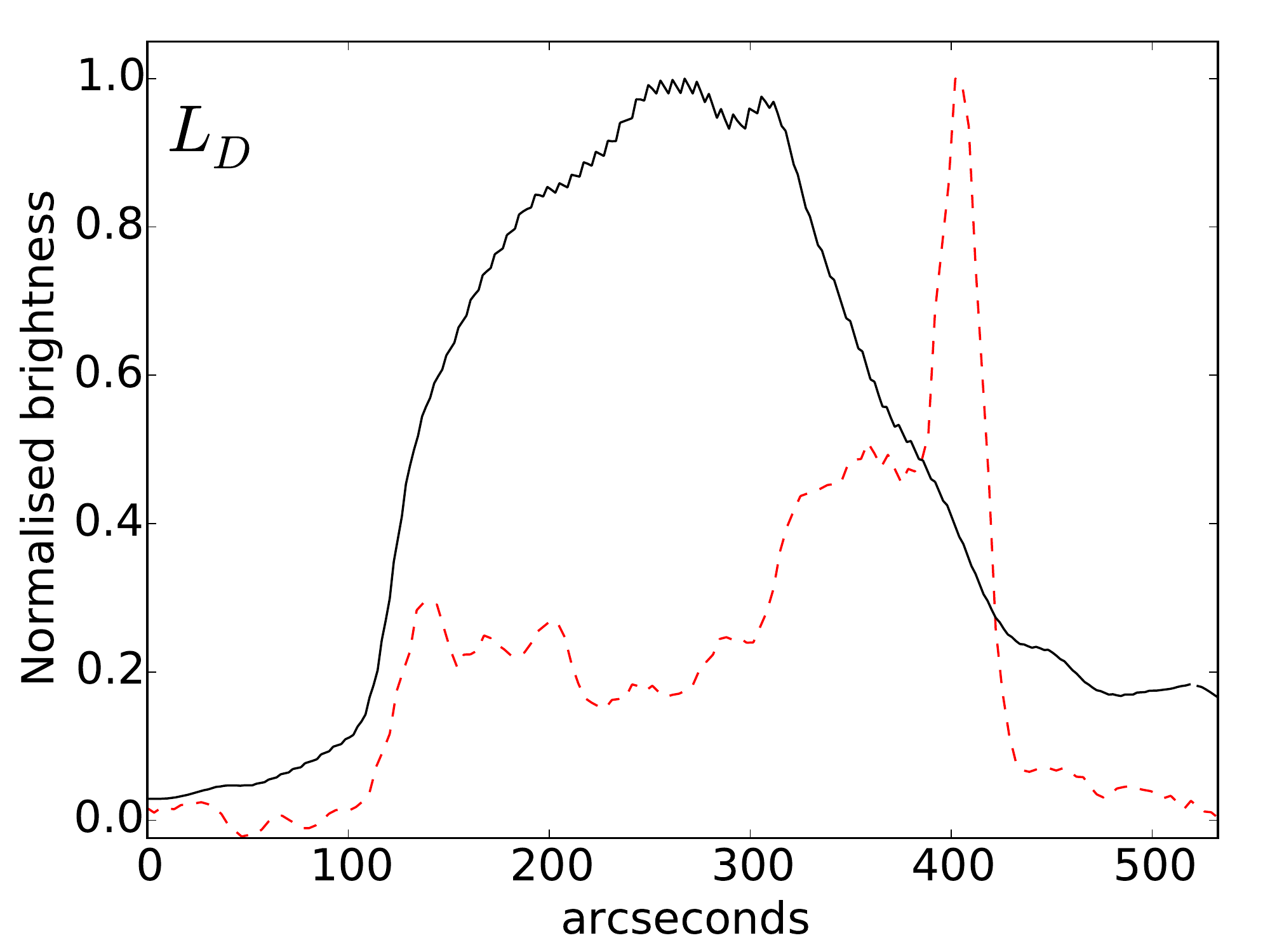}
      \includegraphics[height=4.3cm]{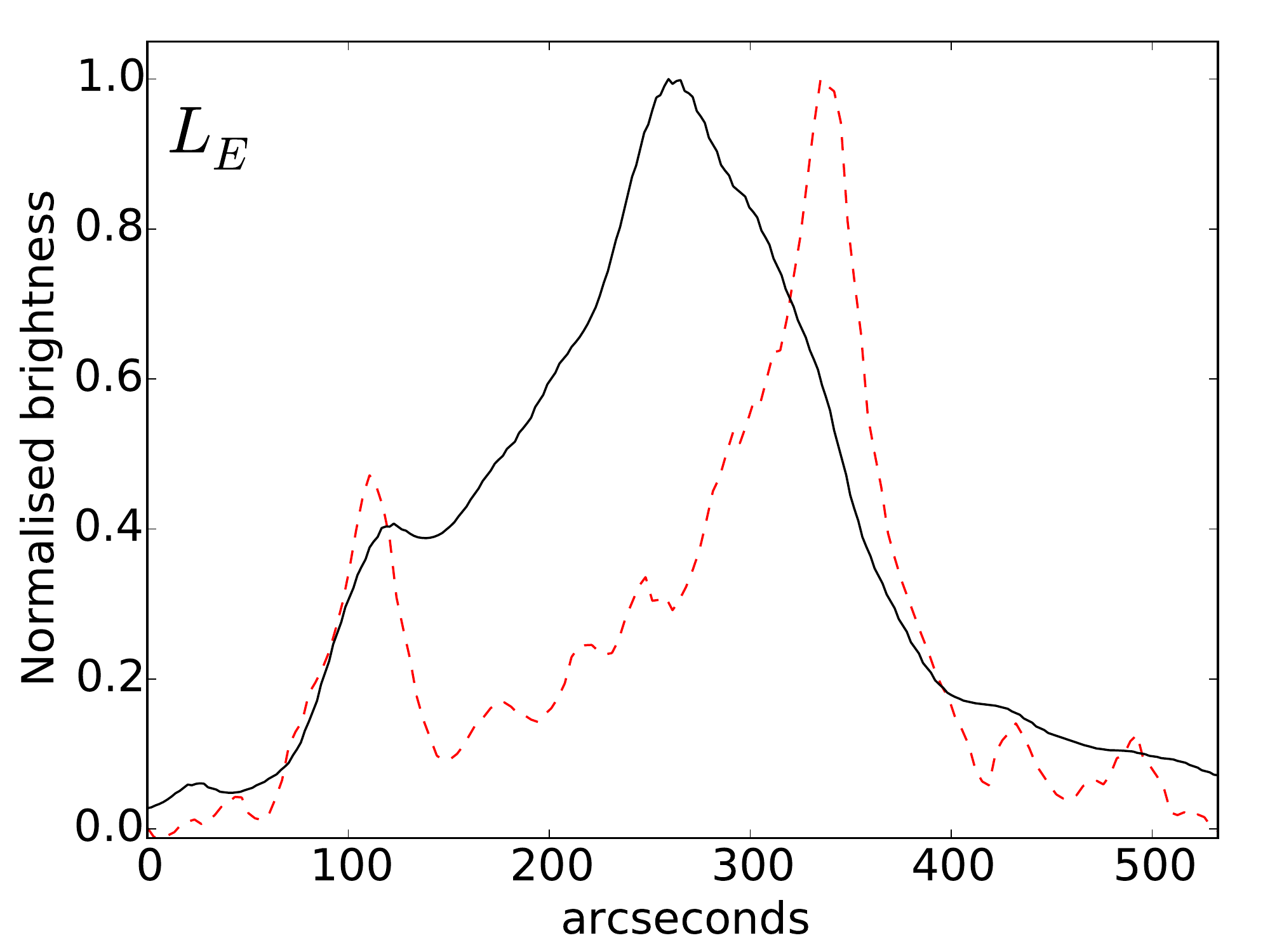}
   \caption{One-dimensional brightness profiles of the X-ray (black lines) and radio emission (red dashed lines). The top left Figure shows the lines $\rm{L_{A}}$ (15:10:30.7 33:32:17.1 to 15:09:55.3 33:33:55.8), $\rm{L_{B}}$ (15:10:30.7 33:31:23.5 to 15:09:55.3 33:33:01.8), $\rm{L_{C}}$ (15:10:30.7 33:29:17.1 to 15:09:55.3 33:30:55.8), $\rm{L_{D}}$ (15:10:18.3 33:34:53.6 to 15:10:05.0 33:26:10.1) and $\rm{L_{E}}$ (15:10:24.3 33:34:53.6 to 15:10:11.0 33:26:10.1) together with the medium-resolution (22$\arcsec \times 19\arcsec$) LOFAR contour image and the high-resolution X-ray intensity image (pixels of size 1.968$\arcsec$)  convolved with a Gaussian kernel of FWHM = 20$\arcsec$ for comparison with the LOFAR image. The LOFAR contours show the $(1,2,4,...)\times 3 \times \sigma_{LOFAR,20\arcsec}$ levels where $\sigma_{LOFAR,20\arcsec} = 350\mu$Jy/beam and the region shown is identical to that presented in Figure \ref{fig:lofar_xray}. The brightness profiles of these images through the 3.3$\arcsec$ radio pixels or 2.0$\arcsec$ X-ray pixels that are crossed by the five profile lines are shown in the other panels as indicated in the top left hand corner of each panel.}
   \label{fig:lofar_brightness_profiles}
\end{figure*}

Brightness profiles through comparable resolution ($\sim20\arcsec$) radio and X-ray images along the likely cluster merging axis ($\sim16^\circ$ from north-south), and approximately perpendicular to this, are shown in Figure \ref{fig:lofar_brightness_profiles}. Where the precise locations of the profiles were selected to avoid contaminating sources in the X-ray and radio images. In the direction of the cluster merger the brightness profiles ($\rm{L_{D}}$ and $\rm{L_{E}}$) show emission that covers the region from the shock front down to region {\it{D}}. The excess of emission in the north of the cluster at the shock front and the bright filaments of emission (region {\it{D}}) dominate the profiles along the direction of the cluster merging axis.  In the perpendicular direction the three profiles we have taken show very different behaviour. In the southern most profile (profile $\rm{L_{C}}$) the radio emission has a similar structure to the X-ray emission but it is narrower and there is a lack of radio emission on the western side. In our middle profile (profile $\rm{L_{B}}$) the radio and X-ray emission have similar extents but there is a large dip in brightness ($\sim 50\%$) in the centre of the radio halo (see Figure \ref{fig:lofar_regions}). The northernmost profile (profile $\rm{L_{A}}$) is across the excess of radio emission that is coincident with the X-ray detected shock front where a peak of steep spectrum radio emission coincident with a cluster member (see Figure \ref{fig:lofar_optical}) is the brightest region of emission. 

\subsection{Diffuse source F}

Source {\it{F}} is a previously undetected, faint diffuse structure at 15:10:05 33:18:59 which lies 620$\arcsec$ (1.86\,Mpc) south of the southern BCG. It has an integrated flux density of $145\pm22$\,mJy at 150\,MHz (within the 4$\sigma_{LOFAR,30\arcsec}$) and an extension along its major axis of 600\,kpc (290$\arcsec$) and 165\,kpc (80$\arcsec$) along its minor axis. Within the same region of the WSRT image of the same resolution, we measure a flux of 1.2$\pm0.3$\,mJy which is a marginal detection given the high noise due to the large primary beam attenuation at this distance from the pointing centre ($12\arcmin$) and would need confirmation. These measurements imply that this object has a spectral index $\alpha \leq -2.15$. Assuming a spectral index of -2.15 we calculate a 1.4\,GHz radio power of $\approx 4.0 \times 10^{22}$\,W\,Hz$^{-1}$ with our adopted cosmology. This steep spectrum object has no obvious optical counterpart (see Figure \ref{fig:lofar_optical}) and at this distance from the cluster centre there is no significant X-ray signal, although it is partially outside of the X-ray field of view. With a peripheral location, lack of optical counterparts and a steep spectrum, this object has the properties expected for a radio relic, but resolved spectral index images and polarisation measurements of this source would help to classify it.

\section{Discussion}

The complexity of the emission seen in Abell 2034 is similar to what is now being uncovered in high-resolution observations of massive merging galaxy clusters that are sufficiently sensitive to radio emission from the ICM (e.g \citealt{Owen_2014}). Whilst good optical, radio and X-ray auxiliary datasets have been helpful to interpret the large variety of complex diffuse radio emission we have observed in Abell 2034, unfortunately many features still can not be classified with certainty, and much of our interpretation remains speculative. A more thorough and robust interpretation and classification of all the diffuse objects we have detected would be achieved with high quality resolved spectral index and polarimetric measurements.

In \cite{Owers_2014} the dynamics of this cluster were described and we outline the key points here to aid the interpretation of the radio emission associated with the ICM. Abell 2034 is primarily a two body head-on collision between a secondary cluster (associated with BCG2) and a primary cluster (associated with BCG1), with the merging axis aligned in the north-south direction and within 23$^\circ$ of the plane of the sky. A secondary subcluster fell into the primary cluster from the south leaving behind an excess of X-ray emission as it was stripped of gas during the infall. During this event the force exerted by the infalling cluster temporarily displaced the X-ray emitting ICM from BCG1 (which is coincident with a peak in the optical galaxy surface density) by 172\,kpc in the north-south direction. The core of the secondary infalling cluster passed the core of the primary cluster $\sim$0.3\,Gyr ago -- a similar time period to the merging events in dissociative mergers such as the  Bullet cluster (\citealt{Markevitch_2002}) and Abell 2146 (\citealt{Russell_2012}). After core passage the secondary subcluster has continued to move north (away from the primary cluster) and has become increasing disrupted. Any cool core that it may have had is now difficult to discern and the shock of low Mach number ($M=1.59^{+0.06}_{-0.07}$), coincident with the galaxies from the subcluster and preceding its gas, is thought to be weakening and slowing down. It is possible that there is also a third substructure involved in the merger and that this is responsible for the X-ray excess in the southern region (\citealt{Kempner_2003}). Indeed, in addition to the large mass concentrations at the positions of two prominent BCGs, \cite{Owers_2014} did observe a local peak in the galaxy surface density $\approx$ 685\,kpc south west of BCG2 coincident with a bright galaxy (see Figure \ref{fig:lofar_optical} or see Figure 7 of \citealt{Owers_2014}). Furthermore, a weak-lensing study with Subaru suggested that Abell 2034 may belong to a large-scale filamentary structure which would further complicate the dynamical properties (\citealt{Okabe_2008}).

\subsection{Diffuse emission in the central region}

Region {\it{E}} hosts the X-ray intensity peak (Figure \ref{fig:lofar_xray}), the detected SZ signal (\citealt{AMI_Planck_2013}) and the highest projected surface mass density (e.g. \citealt{Okabe_2008}). An excess of radio emission close to the shock front was previously observed in this region but its classification was uncertain as it was not detected at high significance.  \cite{Giovannini_2009} suggested it was an irregular elongated radio halo whilst \cite{vanWeeren_2011} noted a poor correspondence with the X-ray emission and suggested it may be a radio relic. Our images confirm the existence of a radio halo in region {\it{E}} of  Abell 2034 which, according to e.g. \cite{Brunetti_2001} and \cite{Petrosian_2001}, suggests that the ongoing merger has introduced turbulence throughout a large area of the ICM where particles have been accelerated sufficiently to produce observable synchrotron radiation. The halo is well characterised in our lower resolution images (see Figure \ref{fig:lofar_xray}) and we find it is most extended in the merging direction (375$\arcsec$ or 775\,kpc in this direction and 280$\arcsec$ or 570\,kpc in the perpendicular direction) similar to what is seen in radio halos in other well characterised merging clusters (e.g. \citealt{Macario_2011},  \citealt{Lindner_2014} and \citealt{Shimwell_2014}) and independently suggesting that the merging is occurring primarily in the north-south direction. However, the structure of the observed radio halo and the diffuse emission embedded within it is very unusual. There is a ridge of enhanced brightness in the north of the halo; a dip in brightness in the centre; a lack of emission on the western side; and an indication of a steep integrated spectral index. We discuss these features in this order in the following paragraphs.

In the northern region of the halo close to the shock front some of the excess brightness at 150\,MHz is due to two bright spots, the brightest being at 15:10:11 +33:33:24, with another fainter structure at 15:10:19 33:32:48. The brightest spot is 45$\arcsec$ south east of BCG2 (this BCG is faint at 150\,MHz but visible at 1.4\,GHz) and is elongated in the direction towards that BCG, it has a very steep spectrum ($\approx-2$) and has a cluster member located within its region of high brightness. The fainter bright spot is slightly extended in the north-west direction and has no clear optical counterpart. There is an optical source close to the brightest region, but it is not certain if it is a cluster member as the redshift was not measured by \cite{Owers_2014}. In addition to these two bright spots that could be associated with cluster members, there is an excess of 150\,MHz radio emission close to the shock -- in the brightness slices we present in Figure \ref{fig:lofar_brightness_profiles}, we avoided the two bright spots and still see a clear brightness ridge in the north. The brightness edge is quite different from what is seen in the Bullet cluster, where a much stronger shock (M=3.0$\pm$0.4) does not produce a relic or enhanced brightness but does cause a very distinctive shape to the edge of the radio halo that traces the shock (\citealt{Shimwell_2014}). However, enhanced brightness is also seen at the position of the southern shock in the `Toothbrush' cluster (1RXS J0603.3+4214): \cite{vanWeeren_2012} suggested this may be a radio relic, but later measured it to have a rather uniform spectral index across its structure, which argues against that interpretation (\citealt{vanWeeren_2016b}). Unfortunately, it is difficult to accurately assess the resolved spectral properties in the region of the Abell 2034 shock as the complete structure is not detected due to the low signal to noise in the 610\,MHz and 1.4\,GHz images and there is contamination due to the two bright spots that were previously discussed. In agreement with  \cite{Owers_2014}, we suggest that for this very weak shock (M=$1.59^{+0.06}_{-0.07}$) in Abell 2034 to produce enhanced emission, there must be a pre-existing population of electrons in this region as a significant amount of direct acceleration from a thermal pool of electrons would require a stronger shock (e.g. \citealt{Hoeft_2007}). If we consider the scenario where the seed relativistic electrons have been very recently reaccelerated by the shock, and they have not had time to cool, then the Mach number can be related to the energy spectrum of the emitting electrons (\citealt{Blandford_1987}) as
\begin{equation}
\delta_{inj} = 2 \frac{M^2 + 1}{M^2-1},
\end{equation}
where $\delta_{inj}$ is the injection power law index of the energy spectrum of emitting electrons, and is related to the injection spectral index by $-\alpha_{inj} = (\delta_{inj}-1)/2$. Hence from the X-ray measured Mach number we may expect a radio spectral index of $\sim$-1.8 which is consistent with our measurements in this region. The pre-existing  electrons may be associated with the two bright spots that we observe close to the shock. The connection between these bright spots and the shock would also explain their apparent elongations in the approximate direction of the shock front.  It would be interesting to know the polarimetric properties of this region and see if it is unpolarised (as expected for a halo) or strongly polarised with polarisation vectors aligned with the shock direction (as expected for a relic).

South of this region of enhanced emission and towards the centre of the radio halo there is a decline ($\sim$50\% in our 20$\arcsec$ resolution images) in the 150\,MHz surface brightness of the radio halo, which is also visible in the 1.4\,GHz images. We have not found any similar decrements in other radio halos in the literature. It could be that the radio halo is concentrated just around the southern region near BCG1 (which is the larger cluster) and we are only seeing emission associated with the shock in the north. In this scenario the passage of the shock could also be responsible for the turbulence which causes the radio halo. A similar explanation was provided for the region of very steep spectral index radio emission between the radio relic and the radio halo in the `Toothbrush cluster' (\citealt{vanWeeren_2016b}), where the size of the region with little radio emission (or the very steep spectrum region) is dictated by the time it takes for the large scale turbulence induced by the passage of the shock to decay into smaller scales for particle acceleration (see \citealt{Brunetti_2007} for details on turbulent decay and acceleration). Assuming this scenario is correct, and that the shock has a constant downstream velocity of 1125\,km/s (where the downstream velocity is the shock velocity of 2057\,km/s divided by the shock compression ratio of 1.83; \citealt{Owers_2014}), then from the $\approx2\arcmin$ (250\,kpc) diameter of the region of decreased brightness in the centre of the halo we can approximate that it takes $\sim$0.22\,Gyr for the large scale turbulence induced by the shock to both decay to smaller scales for particle acceleration and to accelerate electrons to produce observable synchrotron emission.

The brightest region of the radio halo is around the southern BCG and is offset from the peak of the X-ray emission. This offset is best illustrated in Figure \ref{fig:lofar_halo_zoom} or slice $\rm{L_{C}}$ in Figure \ref{fig:lofar_brightness_profiles}, which shows the lack of radio emission in the west of the cluster. The offset between BCG1 (where the peak in the mass distribution is) and the X-ray peak was explained by \cite{Owers_2014} as being due to collisional effects and the  force exerted by the infalling cluster temporarily displacing the X-ray emitting gas. Since the radio emission is similarly displaced, we suggest that the merging is causing a stretching of the ICM and the magnetic field, and that this produces a large amount of turbulence in the region between the BCG and the X-ray peak.

Our spectral index maps do not constrain the spectral index of the radio halo well but our integrated flux measurements at 1.4\,GHz and 150\,MHz suggest that it has a steep spectral index with $\alpha \approx < -1.6$. Whilst more sensitive, higher frequency radio observations are needed to confirm this value and to further characterise the spatial variations, such steep spectra radio halos are expected to be naturally produced in the context of the turbulent re-acceleration scenario (e.g. \citealt{Cassano_2006} and \citealt{Brunetti_2008}) as a consequence of less energetic merger events (mergers between clusters of small mass). With a mass of $\rm{M}_{500}=4.19\pm0.4 \times 10^{14}M_{\odot}$ (\citealt{Planck_2015}), Abell 2034 is one of the least massive cluster systems known to host a radio halo (see e.g. \citealt{Cassano_2013}), so it may be that the energetics involved in the merger are relatively low and this can explain the steep spectrum nature of its emission.  Alternatively, it is also thought that a steep spectral index should be observed during the swich-on or swich-off phase of the radio halo (see e.g. \citealt{Donnert_2013}), and although these periods are short, Abell 2034 is a young merger.  

\subsection{Filamentary emission in the central region}

\begin{figure}
   \centering
   \includegraphics[width=8cm]{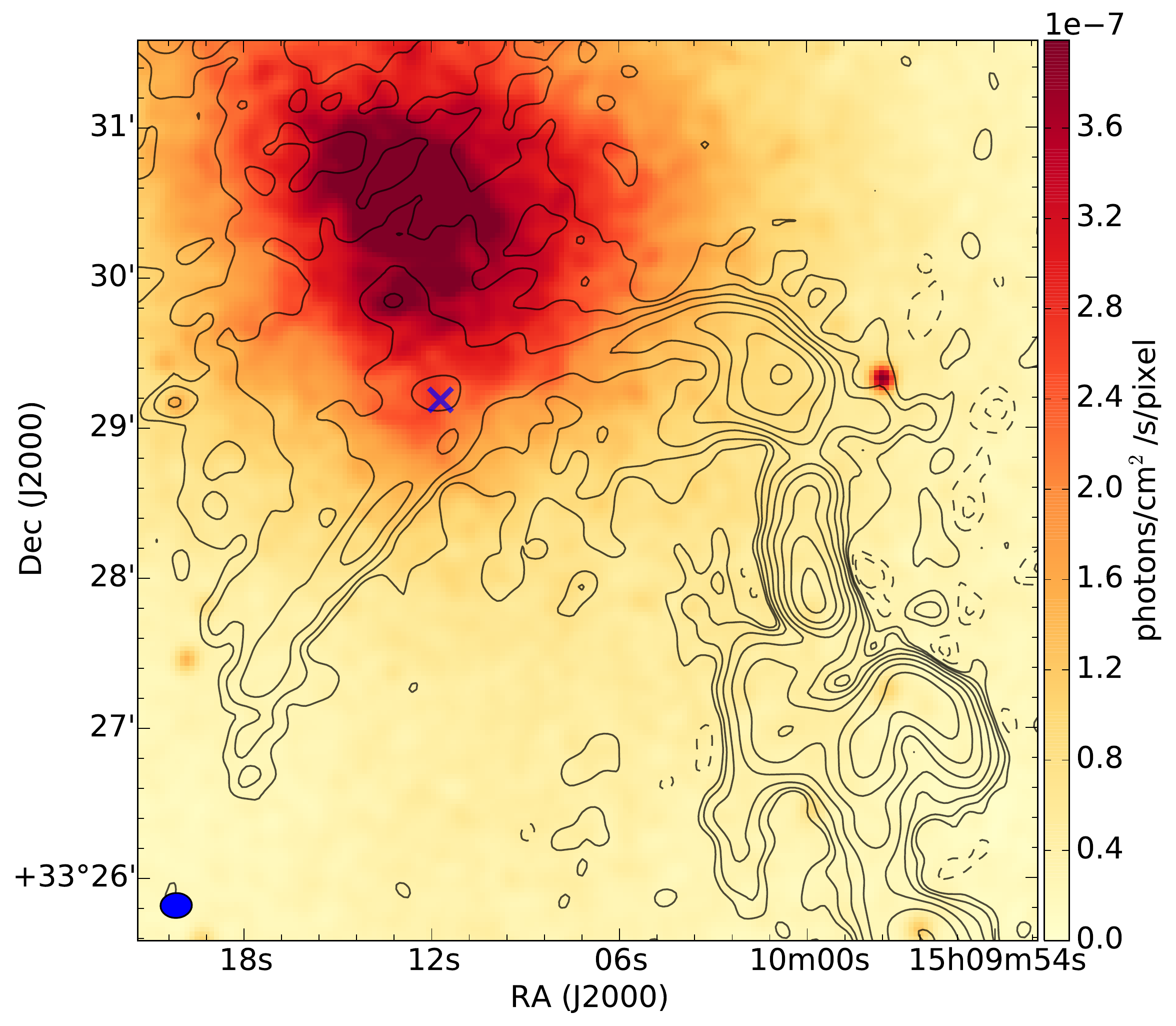}
   \caption{
Contours from a medium-resolution (12$\arcsec \times$ 10$\arcsec$) LOFAR image of Abell 2034 overlaid on a high-resolution \textit{Chandra} image (pixels of 1.968$\arcsec$) which has been smoothed with a Gaussian kernel of FWHM = 6$\arcsec$. The  \textit{Chandra} image is the same as that in Figure \ref{fig:lofar_xray} and the radio image is the same as that in Figure \ref{fig:lofar_regions}. The cross shows the position of BCG1.}
   \label{fig:lofar_halo_zoom}
\end{figure}

Just south of the brightest region of the halo there are narrow filaments of emission which the halo seems to merge into (see Figure \ref{fig:lofar_optical}). This newly discovered filamentary structure has a very steep spectrum and is one of the most intriguing features we have detected in Abell 2034. Coincident with these narrow filaments, there is a rapid decline in the emission of the halo. The lack of radio halo emission south of these narrow filaments gives the halo an asymmetry about its peak in the north-south direction along which the cluster is also merging (see slices $\rm{L_{D}}$ and $\rm{L_{E}}$ of Figure \ref{fig:lofar_brightness_profiles}). A reverse shock with respect to the northern shock seems unlikely to be the cause of this due to the lack of any obvious shock signature in the bright X-ray signal at this position. The shape of the narrow filaments of emission does not trace a path of constant X-ray brightness and the narrow filaments do not have the classic arc-shape seen in e.g. \cite{vanWeeren_2010}. Instead the morphology of the narrow filaments is nearly the reverse of a classic arc  and is similar to that seen in the northern relic of PLCKG287.0 +32.9 (\citealt{Bonafede_2014}). That relic emission was explained as being shock acceleration of a plasma linked to a nearby radio source with typical radio morphology but without an optical counterpart. In Abell 2034, the narrow filaments of emission could again be plasma associated with a nearby galaxy that is now supplying plasma to the radio halo. Whilst there is no obvious connection between cluster galaxies or AGNs and the filaments, BCG1 is just north of the centre of the filamentary structure where there is a break in its direction (see Figures \ref{fig:lofar_optical} and \ref{fig:lofar_headtail}), and the filaments could be associated with fossil plasma from old radio lobes that propagated in the east-west direction away from this BCG. If this fossil plasma were recently compressed by the ongoing merger then its brightness would be enhanced and it may be able to form the well defined structure we observe.  Furthermore, if the narrow filaments of emission are fossil plasma, and were present when the cluster associated with BCG2 fell into BCG1 from the south, it would have disturbed this plasma and spread it northwards explaining the halo emission we see north of the narrow filaments and the lack of emission south of them.  There is a hint of a connection between the central region of the filaments, a region of slightly enhanced X-ray emission close to BCG1 (see Figure \ref{fig:lofar_halo_zoom}) and a region where the X-ray measured temperature is hot and patchy (see Figure 3 of \citealt{Owers_2014}). This may deserve further analysis in the future but it is likely that in this region there is a significant contribution from the hot atmosphere of BCG1. Alternatively, the morphology of the filaments may indicate that they are not associated with the main subcluster, but are seen in projection. It is apparent from X-ray (e.g \citealt{Owers_2014}) and optical studies (e.g. \citealt{Okabe_2008}) that Abell 2034 has very complex dynamics and could possibly be part of a filamentary structure. Whilst there is little evidence that this filamentary structure is along the line of sight, under such complex conditions projection effects can be important and we cannot exclude the possibility that the filamentary structure we see in the radio is a foreground or background relic associated with a subcluster in the Abell 2034 complex and is projected on the X-ray emission from the main cluster. A similar scenario was suggested by \cite{Pizzo_2011} to explain the filamentary radio emission seen close to the centre of Abell 2255.

We have speculated in the previous paragraph that the filamentary structures could be associated with plasma from other radio galaxies (such as BCG1) or a  shock, however, their nature remains very uncertain. The only similar emission that we have found in the literature is the ``Line" in Abell 2256 (see Figure 4 of \citealt{Owen_2014}). In that cluster, the ``Line",  which has no optical counterpart and a very steep spectrum, is detected close to the position of a cold front and a Narrow Angled Tail source (NAT; \citealt{Owen_1976}) but its origin is similarly unclear and its spectral properties have yet to be determined.

\subsection{Diffuse emission in the peripheral regions}
\label{sec:relic_sources}

There are several regions of diffuse radio emission in the peripheral regions of Abell 2034 that we have either discovered or further characterised using our 150\,MHz images. In the following paragraphs we discuss the previously classified radio relic (source  {\it{A}}) and two  newly detected faint and steep spectrum diffuse structures (sources  {\it{B}} and  {\it{F}}).

The previously classified radio relic in region {\it{A}} is bright and compact: with its largest linear size being just 220\,kpc it is one of the smallest radio relics known (e.g. \citealt{Feretti_2012}) and with a Log$(\nu \rm{P}(\nu)/\rm{LLS}^2) = 41.41$\,erg\,s$^{-1}$\,Mpc$^{-2}$ (where LLS is the largest linear size) it has one of the highest surface brightnesses of the known radio relics (see \citealt{Brunetti_2014}). Like most relics, this object is orientated approximately parallel to the closest X-ray brightness edge (see Figure \ref{fig:lofar_xray}) but its location is slightly unusual as it is not in the direction of merging, which is approximately north to south. Unfortunately, in this region  (0.85\,Mpc or 6.8$\arcmin$ from the nearest BCG) there were too few X-ray counts for \cite{Owers_2014} to search for a shock associated with the relic. In our 150\,MHz to 610\,MHz spectral index map (see Figure \ref{fig:lofar_gmrt_spec}) we resolve the minor axis of source {\it{A}} and our measurements indicate that this object does not have the  spectral index gradient that may be expected for a radio relic, with a flatter spectrum on the side furthest from the cluster centre where the electrons have most recently been accelerated. Instead, the resolved spectral index maps suggest a fairly uniform spectral index distribution with no obvious spectral gradients but small variations giving it a patchy appearance. We measure a region of flatter spectral index at the south western tip of the object where there is a faint point-like radio source (see Figure \ref{fig:lofar_optical}). However, this point-like radio emission is likely associated with a massive elliptical galaxy and therefore the correspondence with a faint optical source (shown in Figure 20 of \citealt{vanWeeren_2011}) indicates it is probably at higher redshift than the cluster. Assuming that the diffuse emission in region {\it{A}} is at the redshift of cluster it is possible that it could be associated with an existing population of electrons that has been compressed or reaccelerated. This scenario is similar to what is speculated for the relics shown in e.g. \cite{Bonafede_2014} and \cite{Shimwell_2015}. 
A promising candidate for the source of the existing population of elections is the bright elliptical galaxy at 15:09:40 33:30:40, which \cite{vanWeeren_2011} originally suggested could be responsible for the bright emission from region {\it{A}}.

Just north of source {\it{A}}, there is a very faint, previously unknown diffuse structure (region {\it{B}}) that we cannot robustly classify, but it has similar properties to those expected for a relic or fossil plasma associated with an old AGN.  In our low resolution images, there is a hint of a possible bridge (see Figure \ref{fig:lofar_xray}) that  connects region {\it{B}} to both the north western region of the radio halo and the north of region {\it{A}}. Whilst this bridge is detected at very low significance and would need to be confirmed with deeper observations, it may indicate that there is a physical link between the halo, region {\it{A}} and region {\it{B}}. In this scenario these regions of emission could be one structure that was formed or enhanced by the same shock, albeit one with unusual morphology but complex patterns of shocks are predicted by numerical simulations (e.g. \citealt{Vazza_2009} and \citealt{Skillman_2011}).

Object {\it{F}} is 1.86\,Mpc from the nearest BCG and has a very steep spectral index of at least $\approx$-2.15. Similarly to regions {\it{A}} and {\it{B}} we are unable to robustly classify this source but it again has the properties expected for fossil plasma or a radio relic. Extrapolating its flux to 1.4\,GHz with this spectral index limit indicates that the 1.4\,GHz power of the source is at most $\approx 4.0 \times 10^{22}$\,W\,Hz$^{-1}$ (Log P(1.4\,GHz) = 22.59 W\,Hz$^{-1}$). Therefore, if this object is a radio relic then it is the lowest power radio relic that we are aware of (e.g. \citealt{Feretti_2012}). Furthermore, with a Log$(\nu \rm{P}(\nu)/\rm{LLS}^2) = 39.18$\,erg\,s$^{-1}$\,Mpc$^{-2}$ it would have the lowest surface brightness of a confirmed radio relic  (see Figure 18 of \citealt{Brunetti_2014}). 
At a separation of 1.86\,Mpc in projection ($0.9\times \rm{r_{200}}$ or $1.6\times \rm{r_{500}}$), it is unusually far from the cluster centre although other relics have been detected at comparable distances and several are even further from the X-ray bright region of the cluster (e.g. \citealt{Pizzo_2008} and \citealt{Bonafede_2014}). The interpretation of such distant objects varies, with \cite{Pizzo_2008} favouring a structure formation shock, whilst \cite{Bonafede_2014} suggested the relic could have been created from an earlier core passage between the main merging clusters or an infall of a smaller group into the main cluster.  
It is plausible that object {\it{F}} is a relic caused by a structure formation shock that now has a very steep spectrum as it is fading. Alternatively it could be caused by a third substructure infalling from the south. A third substructure is compatible with X-ray observations as, although \cite{Owers_2014} preferred the two body interpretation for Abell 2034, there is a suggestion from their work and the work of \cite{Kempner_2003} that there is a third smaller subcluster to the south west of BCG1 (see Figure \ref{fig:lofar_xray} or \ref{fig:lofar_optical}). However, assuming this third substructure is associated with Abell 2034, its expected motion is towards or away from the two main components of Abell 2034. This implies that a peripheral shock in the third substructure, on the edge opposite the bulk of the Abell 2034 mass, would likely propagate south west, whereas the putative relic is observed $\approx$600\,kpc directly south of the third substructure.

Finally, we note that the symmetry between the southern relic candidate (region {\it{F}}) and the narrow filaments at the southern part of the radio halo (region {\it{D}}) is difficult to ignore, with object {\it{F}} and the narrow filaments of emission in region {\it{D}} having similar length and curvature. Although, at high-resolution, the structure of the emission in region  {\it{D}} and region  {\it{F}} is different, with the southern relic candidate appearing diffuse whilst the filaments of emission remain narrow and well defined. As indicated by the diamond symbol in Figure \ref{fig:lofar_optical}, almost halfway between the eastern edge of these two objects, at 400\,kpc (195$\arcsec$) from each edge, there is an elliptical galaxy (SDSS J151015.05+332314.4), which is a cluster member and has Mg, Na D and H$\alpha$ absorption features which are expected for a `red and dead' radio galaxy that used to host radio emission but is no longer active (there is no radio emission exactly coincident with this elliptical galaxy --  the closest significant radio point source we detect is displaced by 15$\arcsec$ to the east). Whilst this alignment is most likely a coincidence, as many of the large elliptical galaxies in clusters are `red and dead' with these absorption features, it leads us to suggest that the emission in region {\it{F}} and region {\it{D}} could possibly be the old lobes of a very large radio galaxy where the narrow filaments in region {\it{D}} have been compressed and enhanced by the ongoing merger or revived by a shock. The extent of such a galaxy would be large with the distance from the core to the end of each lobe being $\approx$1\,Mpc.

\subsection{Other unclassified steep spectrum emission}
\label{sec:other_sources}

Besides the candidate relics, the halo and the filamentary structure in Abell 2034, we have discovered further steep spectrum emission. In region {\it{C}}, which is shown in Figure \ref{fig:lofar_headtail}, there is a region of previously unobserved diffuse emission that appears to connect the tailed radio galaxies. Also, north of the $\rm{C_A}$ region there is a bright region ($\rm{D_A}$) without an optical counterpart, whose emission merges with the radio filaments we have briefly discussed in previous sections. These two bright and steep spectrum regions are close to the lobes of tailed radio galaxies and it may be that there are large amounts of aged plasma in these regions which originated from the tailed galaxy lobes and is visible at low frequencies. Or it may be that the emission has been enhanced by reacceleration during the ongoing merging. Similar features are also seen close to the tailed radio galaxies in e.g. the Sausage cluster (\citealt{Stroe_2013}) and Abell 2443 (\citealt{Cohen_2011}). Much of this newly discovered emission could simply be long and complex tails from radio galaxies, where the old plasma is bright at low frequencies and is seen in projection. Although, in this scenario, it is difficult to explain the morphology of source $\rm{D_{A}}$ and its apparent connection to the fainter filaments extending from its northern and southern peripheries which give the impression that $\rm{D_{A}}$  is travelling westwards through the ICM, whereas the nearby tailed galaxy is travelling south.

\section{Conclusions}

With this LOFAR observation we demonstrate the importance of low-frequency observations of merging galaxy clusters and highlight the wealth of additional diffuse steep spectrum sources that can be discovered even in already well studied clusters. Many of the features we have discovered now require resolved spectral index and polarimetric measurements to make robust claims on their origin and to further develop our understanding of the dynamical history of this interesting cluster and the interaction between the ICM and radio galaxies. From our study of Abell 2034 our main conclusions are the following:

\begin{itemize}
\item In region {\it{D}} a bright bulb of emission is connected to two previously undetected filaments of steep spectrum emission which extend across the south of the cluster in the direction perpendicular to the cluster merger axis. We speculate that these may be related to either a shock or the lobes of a radio galaxy which are disturbed or reaccelerated by the ongoing merger. The complication is that neither the bulb or the filaments have an obvious connection with AGNs or cluster galaxies, nor do they correspond to the position of a known X-ray shock. This may be a natural consequence of observing at low frequency, where the sensitivity to steep spectrum emission from old plasma is high.

\item An irregular radio halo (region {\it{E}}) is detected in the region of the highest X-ray intensity. The halo was previously known but LOFAR observations suggest it has a very steep spectrum, with $\alpha < -1.6$. The northern region of the radio halo has an excess of emission close to an X-ray detected shock, as was suggested by previous studies, but it remains uncertain whether this should be classified as a part of the halo, a relic or a combination of the two. The brightness distribution of the halo is complex, we find that there is a decline in the centre and that the region of highest brightness is offset from the BCG. It may be that this complex radio halo structure probes the distribution of turbulence and magnetic fields in the cluster and their interplay with the dynamics of the ICM. Future studies including polarimetry and resolved spectral index analysis are required to investigate this further. 
 \item Three candidate radio relics are detected in the cluster (regions {\it{A}}, {\it{B}} and {\it{F}}).  
 The brightest of the prominent relic candidates is a previously classified relic that lies 0.85\,Mpc west of the cluster and is connected via a very low significance bridge of emission to the northern region of the radio halo. The other prominent relic candidate lies $\approx$2\,Mpc from the cluster centre and has the lowest surface brightness of any previously known relic. It is possible that our observations are beginning to reveal radio emission that is produced by the complex network of shocks which numerical simulations predict will embed dynamically active systems.
\item There are other regions of very steep spectrum emission close to either tailed radio galaxies or the northern shock. Due to its low observing frequency LOFAR is exceptionally sensitive to steep spectrum emission from old plasma and it may be that this emission is related to long and complex tails of radio galaxies seen in projection or the re-acceleration of fossil particles.

\end{itemize}

\section{Acknowledgements}

TS and HR acknowledge support from the ERC Advanced Investigator programme NewClusters 321271. TS and JL thank the Leiden/ESA astrophysics program for summer students (LEAPS) which supported JL in Leiden. M.S.O. acknowledges the funding support from the Australian Research Council through a Future Fellowship Fellowship (FT140100255). CF acknowledges financial support by the ÒAgence Nationale de la RechercheÓ through grant ANR-14-CE23-0004-01. MJH acknowledges support from STFC grant ST/M001008/1. MH acknowledges financial support by the DFG through the Forschergruppe 1254

LOFAR, the Low Frequency Array designed and constructed by ASTRON, has facilities in several countries, that are owned by various parties (each with their own funding sources), and that are collectively operated by the International LOFAR Telescope (ILT) foundation under a joint scientific policy. The National Radio Astronomy Observatory is a facility of the National Science Foundation operated under cooperative agreement by Associated Universities, Inc.

\bsp
\label{lastpage}

\end{document}